\documentclass[12pt,draftcls,onecolumn]{IEEEtran}

\usepackage{amsmath,epsfig}
\usepackage{stmaryrd}
\usepackage{amssymb}
\usepackage{amsfonts}
\usepackage{epic}
\usepackage{graphicx}
\usepackage{curves}
\usepackage{caption}
\usepackage{multirow}
\usepackage{mathrsfs}
\usepackage{pifont}
\usepackage{bbding}
\usepackage{tipa}
\usepackage{bbm}
\usepackage{subfigure}
\usepackage{multicol }
\usepackage{booktabs}
\usepackage{lscape}
\usepackage{amsmath}
\usepackage{amssymb}
\usepackage{latexsym}
\usepackage{multirow}
\usepackage{epsfig}
\usepackage{graphics}
\usepackage{mathrsfs}
\usepackage{algorithm}
\usepackage{algpseudocode}
\usepackage{algorithmicx}
\usepackage{amsmath}

\ifCLASSINFOpdf
\else
\fi

\renewcommand{\baselinestretch}{1.45}
\begin{document}
\newtheorem{theorem}{Theorem}
\newtheorem{proposition}{Proposition}
\newtheorem{definition}{Definition}
\newtheorem{lemma}{Lemma}
\newtheorem{corollary}{Corollary}
\newtheorem{remark}{Remark}
\newtheorem{construction}{Construction}
\newcommand{\supp}{\mathop{\rm supp}}
\newcommand{\sinc}{\mathop{\rm sinc}}
\newcommand{\spann}{\mathop{\rm span}}
\newcommand{\essinf}{\mathop{\rm ess\,inf}}
\newcommand{\esssup}{\mathop{\rm ess\,sup}}
\newcommand{\Lip}{\rm Lip}
\newcommand{\sign}{\mathop{\rm sign}}
\newcommand{\osc}{\mathop{\rm osc}}
\newcommand{\R}{{\mathbb{R}}}
\newcommand{\Z}{{\mathbb{Z}}}
\newcommand{\C}{{\mathbb{C}}}
\newcommand{\tabincell}[3]{\begin{tabular}{@{}#1@{}}#2\end{tabular}}

\title{Power Adaptive Network Coding for a Non-Orthogonal Multiple-Access Relay Channel}

\author{{Sha~Wei, Jun~Li, Wen~Chen, Hang~Su, Zihuai~Lin, and Branka~Vucetic}
\thanks{Sha~Wei, Wen~Chen and Hang~Su are with Department of Electronic Engineering, Shanghai Jiaotong University, Shanghai, China, 200240, email: \{venessa724, wenchen, Hmilyanjohn\}@sjtu.edu.cn.}
\thanks{Jun Li, Zihuai Lin, and Branka Vucetic are with School of Electrical and Information Engineering, The University of Sydney, NSW, 2006, Australia. \{jun.li1,zihuai.lin,branka.vucetic\}@sydney.edu.au}}

\maketitle

%=======================================================================================
\begin{abstract}
In this paper we propose a novel power adapted network coding (PANC) for a non-orthogonal multiple-access relay channel (MARC), where two sources transmit their information simultaneously to the destination with the help of a relay. Different from the conventional XOR-based network coding (CXNC), the relay in our PANC generates network coded bits by considering the coefficients of the source-to-relay channels, and forwards each bit with a pre-optimized power level. Specifically, by defining a symbol pair as two symbols from the two sources, we first derive the exact symbol pair error rate (SPER) of the system. Noting that the generations of the exact SPER are complicated due to the irregularity of the decision regions caused by random channel coefficients, we propose a coordinate transform (CT) method to simplify the derivations of the SPER. Next, we prove that with a power scaling factor at relay, our PANC scheme can achieve full diversity gain, i.e., two-order diversity gain, of the system, while the CXNC can only achieve one-order diversity gain due to multi-user interference. In addition, we optimize the power levels at the relay to equivalently minimize the SPER at the destination concerning the relationship between SPER and minimum Euclidean distance of the received constellation. Simulation results show that (1) the SPER derived based on our CT method can well approximate the exact SPER with a much lower complexity; (2) the PANC scheme with power level optimizations and power scaling factor design can achieve full diversity, and obtain a much higher coding gain than the PANC scheme with randomly chosen power levels.
\end{abstract}

\clearpage
%================================Section_1==================================================
\section{Introduction}
Relaying techniques have been studied for decades to improve the reliability of wireless networks by exploiting spatial diversity via intermediate relay nodes~\cite{Cover:IT76,Laneman:IT04,Liu:book}. Network coding, on the other hand, originated from wire-line networks~\cite{Ahlswede:IT00,Li:IT11}, has been recently leveraged to the wireless networks to enhance the network throughput~\cite{Xor:Sigcomm06}. With the implementations of diversity techniques and network coding at the relay nodes, it is anticipated that the wireless networks can achieve a more reliable communication with a higher network throughput.

Wireless networks can take the advantage of the broadcast nature of wireless signals to further increase the throughput. For instance, by allowing multiple sources transmit in the same channel, less transmissions are needed, and thus a larger network throughput is achieved. However, this kind of non-orthogonal transmissions will lead to multi-user interferences, which could jeopardize the system error performance. Many wireless network coding schemes are designed with the considerations of multi-user interference issue in non-orthogonal transmissions~\cite{Tao:TCOM09,Zhang:JSAC09,Popovski:ICC06,Koike:GLOBE08,Koike:JSAC09}. In~\cite{Tao:TCOM09}, an optimal network coded relay function is derived to minimize the bit error rate in a non-orthogonal two-way relay channel (TWRC). In~\cite{Zhang:JSAC09}, physical-layer network coding is proposed in a non-orthogonal TWRC, where the relay maps the interfered signals from the two sources to a network coded digit. In~\cite{Popovski:ICC06,Koike:GLOBE08,Koike:JSAC09}, denoise-and-forward based network coding schemes are designed for the TWRC with multi-user interferences.

More practical than the TWRC, multiple access relay channel (MARC) has been recognized as a fundamental building block for cellular networks and wireless sensor networks. Different from the TWRC, where perfect side information, i.e., each source's own information, is available at each receiver side, the MARC only has imperfect information at the destination from the sources. The MARC model has attracted a large amount of research interest from both academic and industrial communities~\cite{Kramer:IT05,Bao:TWC08,Ming:TCOM12,Li:VT12,Guan:TWC12,Jun:TWC11,Xiao:TCOM10}. In~\cite{Ming:TCOM12}, Galois field network coding is designed to achieve the full diversity gain of the MARC. In~\cite{Li:VT12}, the authors propose a frame-wise binary filed network coding which enables belief propagation decoding at the destination to achieve the full diversity gain and a high coding gain. However, we note that most of these works consider orthogonal MARC, where multiple sources transmit their signals by using time-division multiple access or frequency-division multiple access manners. Multi-user interferences are not considered in these works, which simplifies the network coding design but leads to a lower spectrum efficiency.

In this paper, we are interested in designing novel network coding schemes for a non-orthogonal MARC over fading channels to achieve the full diversity gain and a good coding gain. Although the conventional CXNC has been shown to achieve full diversity in the orthogonal MARC~\cite{Ming:TCOM12}, we will prove that it cannot achieve full diversity gain in a non-orthogonal one due to multi-user interferences. Specifically in this paper, we consider a two-source, one-relay, one destination non-orthogonal MARC.

There are three major concerns and contributions in our network coding design. First is how to achieve the full diversity gain in the MARC. We propose a novel power adapted network coding (PANC). Different from the CXNC, the relay in our PANC generates network coded bits by considering the coefficients of the source-to-relay channels, and forwards each bit with one of the two given power levels of the relay. Based on the received signals, the relay decides which power level should be applied to each network coded bit. We prove that our PANC scheme with the design of power scaling at relay can achieve the full diversity gain, i.e., two-order diversity gain, of the system, while the CXNC scheme can only achieve one-order diversity gain with or without power scaling design at relay due to multi-user interference.

Secondly, we derive the exact error probability at the destination. By defining a symbol pair as two symbols from the two sources, we develop the exact symbol pair error rate (SPER), with the setup of received constellations at both the relay and the destination. Due to the geometry property of the decision regions, we adopt the wedge probability computation method to investigate the SPER. The wedge probability computation method was first introduced in~\cite{Craig:91} to compute error probability with wedge-shaped decision region in a polar coordinate. Noting that the derivations of the exact SPER are complicated due to the irregularity of the decision regions caused by random channel coefficients, we propose a coordinate transform (CT) method to simplify the derivations. In the CT method, we transform the original parallelogram geometry to a rectangle and approximate the exact results based on original constellation with simple expressions.

Thirdly, we optimize the two power adaption levels at the relay to achieve a higher coding gain. Note that there are quite a few works on the optimizations of the relay transmissions to enhance the error performance over fading channels. In~\cite{Karim:WCL12}, soft information scheme is proposed in fading channels to reduce the error propagation by controlling the transmission power at the relay. However, this scheme is not practical as the optimized power at the relay is a continuous variable. In this paper, we will minimize the SPER by optimizing the two power adaption levels. Specifically, we propose a criteria based on the relationship between Euclidean distance and the SPER, and formulate a convex optimization problem to develop the optimal power adaption levels at the relay.

Simulation results show that (1) the SPER derived based on our CT method can well approximate the exact SPER with a much lower complexity; (2) the PANC scheme with power level optimizations and power scaling factor design can achieve full diversity, and obtain a much higher coding gain than the PANC scheme with randomly chosen power levels; (3) the CXNC scheme cannot achieve full diversity with or without the power scaling design.

The rest of this paper is organized as follows. We first describe the system model in Section II and propose the PANC scheme in Section III. Then we develop the exact system SPER and its approximation in Section IV. In Section V, we address the error propagation problem, and formulate and solve the optimization problem by minimizing the system SPER. Simulation results are presented in Section VI, and conclusions are drawn in Section VII.

The notations used in this work are as follows. $\mathfrak{R}$ and $\mathfrak{I}$ respectively denote the real and the imaginary parts of a complex number. Denote a ray by $l_{ij}$ where the subscript $ij$ is the label of the line.
Denote $AB$ a line segment and $\overline{AB}$ the length between points $A$ and $B$, respectively.
We use the format \emph{ray-vertex-ray} and \emph{ray-vertex-vertex-ray} to describe a wedge and wedge combination, respectively.
Denote $\phi=\angle ABC$ as the crossing angle between line segment $AB$ and $BC$ with the intersection point $B$. The one-dimensional Q-function is defined as $Q_1(x)=\frac{1}{\pi}\int_{0}^{\frac{\pi}{2}}\exp\left(-\frac{x^2}{2\sin^2\theta}\right)d \theta$. The two-dimensional Q-function $Q_2(x,y;\rho)$ with $x=y$ is simplified as $Q_2(x;\rho)$ with expression $Q_2(x;\rho)=\frac{1}{\pi}\int_{0}^{\arctan\left(\sqrt{\frac{1+\rho}{1-\rho}}\right)}\exp\left(-\frac{x^2}{2\sin^2\phi}\right)d\phi$.

%================================================================Section 2=============================================================
\section{System Model}
Consider a two-source single-relay multiple-access relaying system , where two sources $\mathcal{S}_1$ and $\mathcal{S}_2$ transmit their
information to the common destination $\mathcal{D}$ with the assistance of a half-duplex relay node $\mathcal{R}$. Each transmission
period is divided into two transmission phases. In the first transmission phase, the two sources simultaneously broadcast their
symbols $x_1$ and $x_2$ to both the destination and the relay. In the second transmission phase, the two sources keep silent,
while the relay processes the received signals and forwards the network coded symbol $x_{\mathcal{R}}$ to the destination.
At the end of the second phase, the destination decodes the two sources' information based on the received signals.
%\begin{figure}[h]
%\centering
%\includegraphics[width=3.5in]{system_model.eps}
%\caption{The system model of the MARC with two-source, one-relay, and one-destination. The arrows with solid lines represent the first transmission
%phase, and the arrow with a dashed line represents the second transmission phase.}\label{fig:system_model}
%\end{figure}

We assume that all the transmitted signals are binary phase shift keying (BPSK) modulated with equal probability, i.e., $x_1, x_2, x_{\mathcal{R}}\in\{\pm1\}$, and all the signals are transmitted in the same frequency band. The channel between any two given nodes $j$ and $k$, $j \in\{1,2,\mathcal{R}\}, k\in\{\mathcal{R},\mathcal{D}\}$, and $ j\neq k$, is denoted by $h_{jk}$ with the subscripts indicating the nodes under consideration. We assume that $h_{jk}$ for all the $j$ and $k$ are Rayleigh distributed with zero mean and variance $\bar{\gamma}_{jk}$. We consider slow fading channels in our system, i.e., the channels are constant during a transmission period, while independently change from one transmission period to another.

Also, we implement the channel phase pre-equalization for both the source-to-destination multiple access channels (MAC) and the relay-to-destination channel before each transmission. Thus, the effective source-to-destination and relay-to-destination channel coefficients can be regarded as real-valued channels, i.e., real channel coefficients and real values of noise samples.\footnote{Note that, without channel phase pre-equalization, we will encounter a four-dimensional received constellation at destination, namely, the real and imaginary parts of received signal in the first phase, and the real and imaginary parts of received signal in the second phase, respectively, which leave the exact error probability analysis hardly derivable.}

Based on the aforementioned system settings and assumptions, the received signals at the relay and destination in the first transmission phase can be written as
\begin{equation}\label{eq:rec_signal_slot1}
\begin{aligned}
&y_{\mathcal{R}}=\sqrt{E_1}h_{1\mathcal{R}}x_1+\sqrt{E_2}h_{2\mathcal{R}}x_2+n_{\mathcal {R}},\\
&y_1=\sqrt{E_1}|h_{1\mathcal{D}}|x_1+\sqrt{E_2}|h_{2\mathcal{D}}|x_2+n_1,
\end{aligned}
\end{equation}
respectively, where $E_1$ and $E_2$ denote the transmission power of $\mathcal{S}_1$, $\mathcal{S}_2$, respectively, $n_{\mathcal{R}}$ is the complex additive white Gaussian noise (AWGN) sample at the relay with zero mean and variance $\sigma^2/2$ per dimension, and $n_1$ is the real AWGN sample at the destination with zero mean and variance $\sigma^2$.

As we adopt the joint power scaling and adaption scheme at the relay, the instantaneous power at the relay is optimized given the channel realization within each transmission period tending to minimize the SER and achieve full diversity at the destination. Specifically, in the power scaling, we have the scaling factor $\alpha$ ($0\leq \alpha \leq 1$) which is determined based on the channel conditions. In the power adaption, we have two power levels, namely, $a$ and $b$. We have $a^2+b^2\le2E_{\mathcal{R}}^{\text{ave}}$, where $E_{\mathcal{R}}^{\text{ave}}$ is denoted as the average transmission power at the relay. The determinations of $\alpha,a,b$ will be discussed latter in details. Therefore, in the second transmission phase, the received signal at destination can be expressed as
\begin{equation}
y_2=\sqrt{\alpha {E}_{\mathcal{R}}}|h_{\mathcal{R}\mathcal{D}}|x_{\mathcal {R}}+n_2,
\end{equation}
where $n_2$ is the real AWGN at destination with zero mean and variance $\sigma^2$, and ${E}_{\mathcal{R}}\in \{a,b\}$ represents the adaptive power level.

%=================================================================Section 3=============================================================
\section{Network Coded Power Adaption Scheme at Relay}\label{sec:NCPA}
In the conventional CXNC based MARC, XOR operations are implemented at the relay on the two sources' information. We will show later in Section \ref{sec:optimal} that the system cannot achieve full diversity with the conventional network coding. To solve this problem, we propose the PANC scheme, which also take account the source-to-relay channels in network coding design.
 %i.e., based on the received signals, the relay transmits a network coded symbol multiplied with an optimized power level.

Firstly, the relay obtain the two sources' information ($x_1,x_2$) from its received signal $y_{\mathcal{R}}$ by utilizing the maximum likelihood (ML) detection, i.e.,
\begin{equation}\label{eq:detection}
(\hat{x}_1,\hat{x}_2)=\mathop{\arg\min}\limits_{\scriptstyle \tilde{x}_1,\tilde{x}_2 \in \{\pm 1\}\hfill\atop
\scriptstyle \hfill}\bigg|y_{\mathcal{R}}-\sqrt{E_{1\mathcal{R}}}h_{1\mathcal{R}}\tilde{x}_1-\sqrt{E_{2\mathcal{R}}}h_{2\mathcal{R}}\tilde{x}_2\bigg|^2,
\end{equation}
where $(\hat{\cdot})$ denotes the detected symbol, and $(\tilde{\cdot})$ denotes the trial symbol used in the hypothesis-detection problem. Then the relay takes network coding operation on the two detected symbols. The network coded operation in our PANC is denoted by $\boxplus$, which is different from the conventional XOR operation. That is, we calculate $x_{\mathcal{R}}$ by $x_{\mathcal{R}}=\hat{x}_1\boxplus \hat{x}_2=\sign(|h_{1\mathcal{R}}|\hat{x}_1+|h_{2\mathcal{R}}|\hat{x}_2)$. Next, the relay chooses the power level $E_\mathcal{R}$ based on the decoded symbols, i.e., if $(\hat{x}_1=1,\hat{x}_2=1)$ or $(\hat{x}_1=-1,\hat{x}_2=-1)$, power level is chosen as $a$; else if $(\hat{x}_1=1,\hat{x}_2=-1)$ or $(\hat{x}_1=-1,\hat{x}_2=1)$, power level is chosen as $b$. The reason for adopting the new proposed network coding operation and power level allocation method is that the received constellation at destination is a parallelogram, on which the $(\hat{x}_1=1,\hat{x}_2=1)$ corresponding constellation point lies in a diagonal with the $(\hat{x}_1=-1,\hat{x}_2=-1)$ corresponding constellation point. While for XOR operation, the received constellation is an irregular quadrilateral no matter what power level allocation result we implement.

The values of $a$ and $b$ are optimized by minimizing the system SPER at the destination. Suppose that the instantaneous channel state information (CSI) of the MARC is available to the destination before the transmission period starts. The destination first optimizes the power levels $a$ and $b$ by minimizing the instantaneous SPER based on the CSI, and then feedbacks the values to the relay before the transmission period starts. The relay will use this power adaption to transmit the network coded symbol $x_{\mathcal{R}}$. Note that, as the values of $a$ and $b$ is derived based on the instantaneous CSI, their values will keep invariant within each transmission period, and change from one transmission period to another.

The PANC scheme at the relay can also be illustrated by a two-dimensional instantaneous relay constellation (IRC), which is associated with the SPER calculation in the next section. The signal part of $y_{\mathcal{R}}$, i.e., $\sqrt{E_1}h_{1\mathcal{R}}x_1+\sqrt{E_2}h_{2\mathcal{R}}x_2$, can be seen as a point in the IRC with X-axis being its real part, and Y-axis being its imaginary part. We define the constellation points (CPs) $V_i$ of the IRC, $i=1,\cdots,4$, to represent the four possible values of $\pm \sqrt{E_1}h_{1\mathcal{R}}\pm \sqrt{E_2}h_{2\mathcal{R}}$, and define the sources' symbol pairs by $T_i \triangleq (x_1,x_2)$. Specifically, we have $T_1 \triangleq (1,1),T_2 \triangleq (-1,1),T_3 \triangleq (1,-1)$, and $T_4 \triangleq (-1,-1)$.

%The relationship between the CP $V_i$ and the source symbol pair $T_i$ is shown in Table~\ref{table:coordinate}, where $\Re\{\cdot\}$ and $\Im\{\cdot\}$ represent the real and imaginary parts of a complex number, respectively.

\begin{figure}[h]
\begin{center}
\vspace{-5ex}
\includegraphics[width=3.5in]{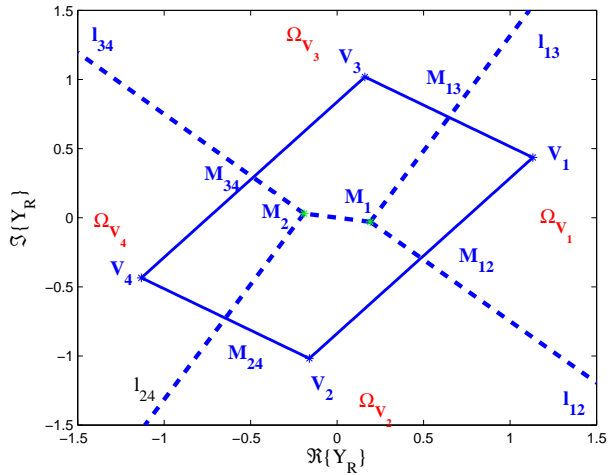}
\caption{One possible instantaneous relay constellation, where dashed lines represent boundaries of decision regions.}
\label{figure_relay_no}
\vspace{-5ex}
\end{center}
\end{figure}

From Fig.~\ref{figure_relay_no}, we can see that the geometry of the IRC composed by CPs $V_i$ is a parallelogram. Similar to Voronoi diagram in~\cite{multi_user}, the decision regions are segmented by the perpendicular bisectors of each side of the parallelogram. Specifically, rays $l_{12}$, $l_{13}$, $l_{24}$ and $l_{34}$ are perpendicular bisectors of sides $V_1V_2$, $V_1V_3$, $V_2V_4$ and $V_3V_4$, respectively, $M_1$ is the crossing points of rays $l_{12}$ and $l_{13}$, and $M_2$ is the crossing points of rays $l_{24}$ and $l_{34}$. $M_{ij}$ is the middle point of each side $V_iV_j$. The decision region $\Omega_{V_1}$ of $V_1$, defined as wedge $l_{12}-M_1-l_{13}$ in Fig~\ref{figure_relay_no}, is given by
\begin{equation}
\begin{aligned}
\Omega_{V_1}\triangleq &\left\{\frac{\Re\{h_{2\mathcal{R}}\}}{\Im\{h_{2\mathcal{R}}\}}\Re\{y_{\mathcal{R}}\}+\Im\{y_{\mathcal{R}}\}-\Im\{h_{1\mathcal{R}}\}
-\frac{\sqrt{E_1}\Re\{h_{1\mathcal{R}}\}\Re\{h_{2\mathcal{R}}\}}{\Im\{h_{2\mathcal{R}}\}}<0 \right. \cap\\
&\left.
\frac{\Re\{h_{1\mathcal{R}}\}}{\Im\{h_{1\mathcal{R}}\}}\Re\{y_{\mathcal{R}}\}+\Im\{y_{\mathcal{R}}\}-\Im\{h_{2\mathcal{R}}\}
-\frac{\sqrt{E_2}\Re\{h_{1\mathcal{R}}\}\Re\{h_{2\mathcal{R}}\}}{\Im\{h_{1\mathcal{R}}\}}\geq0\right\}.
\end{aligned}
\end{equation}

Similarly, we can obtain the decision regions of $V_2$, $V_3$ and $V_4$, denoted by $\Omega_{V_2}$, $\Omega_{V_3}$, and $\Omega_{V_4}$, respectively. Based on the four decision regions, we have the one-to-one mapping between the CPs and the network coded power level as
\begin{equation}
\sqrt{E_{\mathcal{R}}}x_{\mathcal{R}}=\left\{\begin{array}{cc}
a & \hspace{2mm}\text{if}\hspace{2mm} (\Re\{y_{\mathcal{R}}\},\Im\{y_{\mathcal{R}}\}) \in \Omega_{V_1},\\
b & \hspace{2mm}\text{if}\hspace{2mm} (\Re\{y_{\mathcal{R}}\},\Im\{y_{\mathcal{R}}\}) \in \Omega_{V_2},\\
-b & \hspace{2mm}\text{if}\hspace{2mm} (\Re\{y_{\mathcal{R}}\},\Im\{y_{\mathcal{R}}\}) \in \Omega_{V_3},\\
-a & \hspace{2mm}\text{if}\hspace{2mm} (\Re\{y_{\mathcal{R}}\},\Im\{y_{\mathcal{R}}\}) \in \Omega_{V_4},
\end{array}
\right.
\end{equation}

Based on the observations $y_1$ and $y_2$, the destination jointly decode the two source symbols with the minimum Euclidean distance detection. Then we have
\begin{equation}\label{eq:mdd}
(\hat{x}_1,\hat{x}_2)=\mathop {\arg \min }\limits_{\scriptstyle \tilde{x}_1,\tilde{x}_2 \in \{\pm 1\}\hfill\atop
\scriptstyle \hfill}\left(
\bigg| y_1-\sum_{j=1}^2|h_{j\mathcal{D}}|\tilde{x}_j
\bigg|^2+\bigg|y_2-|h_{\mathcal{RD}}|\sqrt{\tilde{E}_{\mathcal{R}}}\left(\tilde{x}_1\boxplus \tilde{x}_2\right)
\bigg|^2\right),
\end{equation}
where $\tilde{E}_{\mathcal{R}}\in\{a,b\}$ is determined by $\tilde{x}_1$ and $\tilde{x}_2$.

%=================================================================Section 4=============================================================
\section{Error Performance Analysis}\label{sec:error}
In this section, we investigate the instantaneous SPER performance of the PANC given a channel realization vector $\textbf{h}=[h_{1\mathcal{R}},h_{2\mathcal{R}},h_{1\mathcal{D}},h_{2\mathcal{D}},h_{\mathcal{RD}}]$. Assuming symbols are transmitted with equal probability, then the general expression of the system SPER of the PANC scheme is given by
\begin{equation}\label{eq:sys}
P_{e,inst}=\sum_{i=1}^4P(\mathcal{E}|T_i,\textbf{h})P(T_i)=\frac1{4}\sum_{i=1}^4P(\mathcal{E}|T_i,\textbf{h}),
\end{equation}
where $T_i$ is the symbol pair defined in Section \ref{sec:NCPA}, $\mathcal{E}$ denotes the symbol error event at the destination that a transmitted symbol pair from two sources is decoded to an erroneous pair, i.e., either $x_1$ or $x_2$ is wrongly detected or both $x_1$ and $x_2$ are wrongly detected, $P(\mathcal{E}|T_i,\textbf{h})$ is the conditional SPER given $T_i$ is transmitted and the channel realization $\textbf{h}$, and $P(T_i)=\frac1{4}$ is the probability that $T_i$ is sent by the two sources. Since the decision regions of $T_1$ and $T_4$ are symmetric, and the decision regions of $T_2$ and $T_3$ are symmetric, we have $P(\mathcal{E}|T_1)=P(\mathcal{E}|T_4)$ and $P(\mathcal{E}|T_2)=P(\mathcal{E}|T_3)$. Therefore, \eqref{eq:sys} can be rewritten as
\begin{equation}\label{eq:general}
\begin{split}
P_{e,inst}&=\frac{1}{2}\left(P(\mathcal{E}|T_1,\textbf{h})+P(\mathcal{E}|T_2,\textbf{h})\right)\\
&=\frac{1}{2}\left\{\sum_{i=1}^2\sum_{k \in \{\pm a, \pm b\}}P(\mathcal{E}|k,T_i,h_{1\mathcal{D}},h_{2\mathcal{D}},h_{\mathcal{RD}})P(\sqrt{E_{\mathcal{R}}}x_{\mathcal{R}}=k|T_i,h_{1\mathcal{R}},h_{2\mathcal{R}})\right\},
\end{split}
\end{equation}
where $P(\sqrt{E_{\mathcal{R}}}x_{\mathcal{R}}=k|T_i,h_{1\mathcal{R}},h_{2\mathcal{R}})$ is the conditional probability that $x_{\mathcal{R}}$ is allocated to the power level $|k|$ at the relay, and $P(\mathcal{E}|k,T_i,h_{1\mathcal{D}},h_{2\mathcal{D}},h_{\mathcal{RD}})$ is the conditional probability that the destination makes a wrong decision on the sources' symbol pair $T_i$. For the sake of simplicity, we omit the channel coefficients in these notations, and use $P(\sqrt{E_{\mathcal{R}}}x_{\mathcal{R}}=k|T_i)$ and $P(\mathcal{E}|k,T_i)$.

In the following we will derive $P_{e,inst}$ based on two methods. The first method is based on wedge probability computation~\cite{Craig:91,Peng:WCOM09}, which derives the instantaneous SPER of the system. However, the calculations of various wedge probabilities are very complicated and time-consuming. Hence, we simplify the wedge probabilities by applying coordinate transformations with a little accuracy loss.

%========================================================================================
\subsection{SPER based on Wedge Probabilities}
We now investigate the SPER performance of the PANC scheme based on the wedge probability computation method. Due to the randomness of channel realizations, the two-dimension decision regions of a symbol pair $T_i$ at both the relay and the destination are irregular and wedge-like, e.g., one possible case of the received constellation and its corresponding decision regions at the relay is shown in Fig.~\ref{figure_relay_no}. We thus use the wedge probability computation method to facilitate the derivation of the SPER. There are totally five basic wedge prototypes, discussed in Appendix~\ref{sub:wedge}. Corresponding to these five wedge prototypes, we derive the corresponding five wedge probabilities in Appendix~\ref{sub:wedge}, i.e., $P_{w_i}$ for $i\in\{1,\cdots,5\}$, based on which, we will later derive the SPER of the system.

We first focus on the probabilities $P(\sqrt{E_{\mathcal{R}}}x_{\mathcal{R}}=k|T_i)$, $k\in\{\pm a, \pm b\}$, at the relay. With the setup of IRC given in Section~\ref{sec:NCPA}, we calculate the probabilities that relay detects the received signal successfully, i.e., $P(\sqrt{E_{\mathcal{R}}}x_{\mathcal{R}}=a|T_1)$ and $P(\sqrt{E_{\mathcal{R}}}x_{\mathcal{R}}=b|T_2)$, by $P_{w_4}$ and $P_{w_5}$ defined in equations~\eqref{eq:wedge_inside1} and~\eqref{eq:wedge_inside2} respectively, and calculate the probabilities that relay makes wrong decisions, i.e., $P(\sqrt{E_{\mathcal{R}}}x_{\mathcal{R}}\in \{\pm b,-a\}|T_1)$ and $P(\sqrt{E_{\mathcal{R}}}x_{\mathcal{R}}\in \{\pm b,-a\}|T_2)$, by $P_{w_1}$, $P_{w_2}$, and $P_{w_3}$ defined in equations \eqref{eq:wedge1}, \eqref{eq:wedge2}, and \eqref{eq:wedge3}, respectively. The computation of exact instantaneous correct probabilities and error probabilities are divided into six cases according to the relative lengths and intersection angles of parallelogram's neighboring sides, and intersection angles of perpendicular bisectors. These six cases are given as
\begin{equation}
\left\{ {\begin{array}{*{20}{c}}
{\overline {{V_1}{V_4}}  > \overline {{V_2}{V_3}} \left\{ {\begin{array}{*{20}{c}}
{{M_1},{M_2} \notin \mathcal{P}\left\{ {\begin{array}{*{20}{c}}
{\overline {{V_1}{V_2}}  > \overline {{V_1}{V_3}} ,\text{Case one}}\\
{\overline {{V_1}{V_2}}  \le \overline {{V_1}{V_3}} ,\text{Case two}}
\end{array}} \right.}\\
{{M_1},{M_2} \in \mathcal{P},{\kern 1pt} {\kern 1pt} {\kern 1pt} \text{Case three}}
\end{array}} \right.}\\
{\overline {{V_1}{V_4}}  \le \overline {{V_2}{V_3}} \left\{ {\begin{array}{*{20}{c}}
{{M_1},{M_2} \notin \mathcal{P}\left\{ {\begin{array}{*{20}{c}}
{\overline {{V_1}{V_2}}  > \overline {{V_1}{V_3}} ,\text{Case four}}\\
{\overline {{V_1}{V_2}}  \le \overline {{V_1}{V_3}} ,\text{Case five}}
\end{array}} \right.}\\
{{M_1},{M_2} \in \mathcal{P},{\kern 1pt} {\kern 1pt} {\kern 1pt} \text{Case six}}
\end{array}} \right.}
\end{array}} \right.
\end{equation}

In \eqref{equ:various_p_at_relay}, we show an example of computing $P(\sqrt{E_{\mathcal{R}}}x_{\mathcal{R}}=k|T_i)$ for Case three, i.e., $\overline{V_1V_4}>\overline{V_2V_3}$ and $M_1,M_2\in\mathcal{P}$, where $\mathcal{P}$ is defined as the region of parallelogram in IRC. Let $d_{ik}$ be the normalized distance between $V_i$ and $M_k$, which is given by $d_{ik}=\frac{|V_i-M_k|^2}{\sigma^2}$. According to the law of total probability, $P(\sqrt{E_{\mathcal{R}}}x_{\mathcal{R}}=-a|T_1)=1-\sum\limits_{k\in \{a,\pm b\}}P(\sqrt{E_{\mathcal{R}}}x_{\mathcal{R}}=k|T_1)$ and $P(\sqrt{E_{\mathcal{R}}}x_{\mathcal{R}}=-a|T_2)=1-\sum\limits_{k\in \{a,\pm b\}}P(\sqrt{E_{\mathcal{R}}}x_{\mathcal{R}}=k|T_2)$. We can obtain the probabilities for the other five cases by following the similar method.

\begin{figure*}[!t]\small{\vspace{-5ex}
\begin{equation}\label{equ:various_p_at_relay}
\small
\begin{aligned}
&P(\sqrt{E_{\mathcal{R}}}x_{\mathcal{R}}=a|T_1)=P_{w_4}\left(d_{11},\arcsin\left(\frac{\overline{V_1M_{13}}}{\overline{V_1M_1}}\right),\arcsin\left(\frac{\overline{V_1M_{12}}}{\overline{V_1M_1}}\right)\right),\\
&P(\sqrt{E_{\mathcal{R}}}x_{\mathcal{R}}=b|T_1)=P_{w_3}(d_{12},d_{11},\angle V_1M_2M_{24},\pi-\angle V_1M_2M_1,\angle
V_1M_1M_{12},\angle V_1M_1M_2,1,1),\\
&P(\sqrt{E_{\mathcal{R}}}x_{\mathcal{R}}=-b|T_1)=P_{w_3}(d_{11},d_{12},\pi-\angle V_1M_1M_2,\pi-\angle
V_1M_1M_{13},\angle V_1M_2M_{1},\angle V_1M_2M_{34},1,1),\\
&P(\sqrt{E_{\mathcal{R}}}x_{\mathcal{R}}=b|T_2)=P_{w_5}\left(d_{21},d_{22},\pi-\arcsin\left(\frac{\overline{V_2M_{12}}}{\overline{V_2M_1}}\right),\arcsin\left(\frac{\overline{V_2M_{24}}}{\overline{V_2M_2}}\right),\angle
M_1V_2M_{2},\angle V_2M_1M_2\right),\\
&P(\sqrt{E_{\mathcal{R}}}x_{\mathcal{R}}=a|T_2)=P_{w_2}(d_{21},\angle V_2M_1M_{12},\pi-\angle V_2M_1M_{13}),\\
&P(\sqrt{E_{\mathcal{R}}}x_{\mathcal{R}}=-a|T_2)=P_{w_2}(d_{21},\angle V_2M_1M_{34},\pi-\angle V_2M_1M_{13}).
\end{aligned}
\end{equation}}
\hrulefill \vspace*{4pt}
\vspace{-5ex}
\end{figure*}

\begin{figure}
\centering
\includegraphics[width=3.5in]{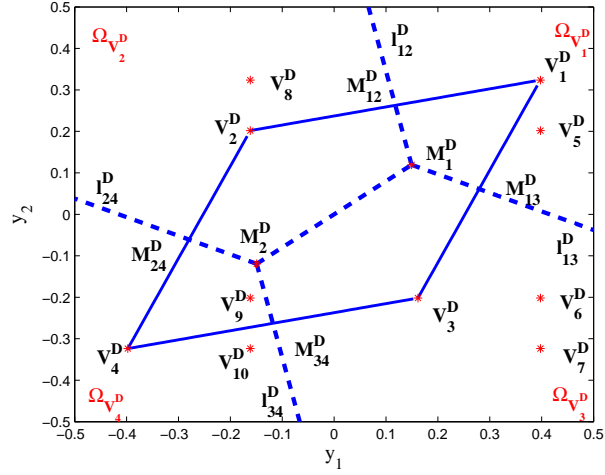}\\
\caption{Instantaneous destination constellation for network coded power adaption scheme. In particular, dashed lines represent boundaries of decision regions $\Omega_{V_i^{\mathcal{D}}}$.}\label{figure_dest}
\vspace{-5ex}
\end{figure}

Then we consider the conditional error probabilities, $P(\mathcal{E}|k,T_i)$, at the destination. we establish an \emph{instantaneous destination constellation} (IDC) with X-axis being $y_1$ and Y-axis being $y_2$ based on the minimum Euclidean distance detection, as shown in Fig.~\ref{figure_dest}. When the relay detects the received signal successfully and forwards the correct symbol, we define four reference points as \begin{equation}\label{eq:correct_RP_dest}
\begin{aligned}
V_1^{\mathcal{D}}&=(\sqrt{E_1}|h_{1\mathcal{D}}|+\sqrt{E_2}|h_{2\mathcal{D}}|,a|h_{\mathcal{RD}}|),\hspace{2mm}
V_2^{\mathcal{D}}=(-\sqrt{E_1}|h_{1\mathcal{D}}|+\sqrt{E_2}|h_{2\mathcal{D}}|,b|h_{\mathcal{RD}}|),\\
V_3^{\mathcal{D}}&=(\sqrt{E_1}|h_{1\mathcal{D}}|-\sqrt{E_2}|h_{2\mathcal{D}}|,-b|h_{\mathcal{RD}}|),\hspace{2mm}
V_4^{\mathcal{D}}=(-\sqrt{E_1}|h_{1\mathcal{D}}|-\sqrt{E_2}|h_{2\mathcal{D}}|,-a|h_{\mathcal{RD}}|).
\end{aligned}
\end{equation}

Similar to IRC, the decision regions of IDC are segmented by the perpendicular bisectors of each edge in parallelogram according to Voronoi rule \cite{multi_user}. In particular, rays $l_{12}^{\mathcal{D}}$, $l_{13}^{\mathcal{D}}$, $l_{24}^{\mathcal{D}}$ and $l_{34}^{\mathcal{D}}$ are perpendicular bisectors of sides $V_1^{\mathcal{D}}V_2^{\mathcal{D}}$, $V_1^{\mathcal{D}}V_3^{\mathcal{D}}$, $V_2^{\mathcal{D}}V_4^{\mathcal{D}}$ and $V_3^{\mathcal{D}}V_4^{\mathcal{D}}$, respectively. $M_1^{\mathcal{D}}$ is the crossing points of ray $l_{12}^{\mathcal{D}}$ and $l_{13}^{\mathcal{D}}$, and $M_2^{\mathcal{D}}$ is the crossing points of ray $l_{24}^{\mathcal{D}}$ and $l_{34}^{\mathcal{D}}$. Line segment $\overline{M_1^{\mathcal{D}}M_2^{\mathcal{D}}}$ is the perpendicular bisector of diagonal $\overline{V_2^{\mathcal{D}}V_3^{\mathcal{D}}}$. Then the correct decision region $V_1^{\mathcal{D}}$ at destination is
\begin{equation}
\Omega_{V_1^{\mathcal{D}}}\triangleq\left\{\frac{2\sqrt{E_1}|h_{1\mathcal{D}}|}{(a-b)|h_{\mathcal{RD}}|}y_1+y_2+M_1^{\mathcal{D}}>0
\bigcap \frac{2\sqrt{E_2}|h_{2\mathcal{D}}|}{(a+b)|h_{\mathcal{RD}}|}y_1+y_2+M_2^{\mathcal{D}}>0
\right\},
\end{equation}
where $M_1^{\mathcal{D}}=-\frac{1}{2}(a+b)|h_{\mathcal{RD}}|-2\sqrt{E_1E_2}|h_{1\mathcal{D}}||h_{2\mathcal{D}}|(a-b)|h_{\mathcal{RD}}|$ and $M_2^{\mathcal{D}}=-\frac{1}{2}(a-b)|h_{\mathcal{RD}}|-2\sqrt{E_1E_2}|h_{1\mathcal{D}}||h_{2\mathcal{D}}|(a+b)|h_{\mathcal{RD}}|$. Likewise, we can obtain $\Omega_{V_i^{\mathcal{D}}}$ for $i=2,3,4$.

When $T_1$ and $T_2$ are transmitted by the sources, the error probabilities at the destination given that relay transmits the correct symbol is denoted by $P\left(\mathcal{E}|\sqrt{E_{\mathcal{R}}}x_{\mathcal{R}}=a,T_1\right)$ and $P\left(\mathcal{E}|\sqrt{E_{\mathcal{R}}}x_{\mathcal{R}}=b,T_2\right)$, respectively. Let $d_{ik}^{\mathcal{D}}$ be the normalized distance between $V_i^{\mathcal{D}}$ and $M_k^{\mathcal{D}}$ in Fig.~\ref{figure_dest}, which is given by $d_{ik}^{\mathcal{D}}=\frac{|V_i^{\mathcal{D}}-M_k^{\mathcal{D}}|^2}{\sigma^2}$. Regarding the probability $P\left(\mathcal{E}|\sqrt{E_{\mathcal{R}}}x_{\mathcal{R}}=a,T_1\right)$, we have
\begin{equation}\label{eq:perfect1}
P\left(\mathcal{E}|\sqrt{E_{\mathcal{R}}}x_{\mathcal{R}}=a,T_1\right)=1-P_{w_4}\left(d_{11}^{\mathcal{D}},\phi_1,\phi_2\right).
\end{equation}
Defining $\mathcal{P}^\mathcal{D}$ as the region of parallelogram in IDC, the determinations of $\phi_1,\phi_2$ in \eqref{eq:perfect1} are as follows. When $\overline{V_1^{\mathcal{D}}V_2^{\mathcal{D}}}<\overline{V_1^{\mathcal{D}}V_3^{\mathcal{D}}}$ and $M_1^{\mathcal{D}},M_2^{\mathcal{D}}\in\mathcal{P}^{\mathcal{D}}$, we have $\phi_1=\pi-\arcsin\left(\frac{\overline{V_1^{\mathcal{D}}M_{13}^{\mathcal{D}}}}{\overline{V_1^{\mathcal{D}}M_1^{\mathcal{D}}}}\right)$, $\phi_2=\arcsin\left(\frac{\overline{V_1^{\mathcal{D}}M_{12}^{\mathcal{D}}}}{\overline{V_1^{\mathcal{D}}M_1^{\mathcal{D}}}}\right)$; When $\overline{V_1^{\mathcal{D}}V_2^{\mathcal{D}}}\geq\overline{V_1^{\mathcal{D}}V_3^{\mathcal{D}}}$ and $M_1^{\mathcal{D}},M_2^{\mathcal{D}}\not\in\mathcal{P}^{\mathcal{D}}$, we have $\phi_1=\arcsin\left(\frac{\overline{V_1^{\mathcal{D}}M_{13}^{\mathcal{D}}}}{\overline{V_1^{\mathcal{D}}M_1^{\mathcal{D}}}}\right)$, $\phi_2=\pi-\arcsin\left(\frac{\overline{V_1^{\mathcal{D}}M_{12}^{\mathcal{D}}}}{\overline{V_1^{\mathcal{D}}M_1^{\mathcal{D}}}}\right)$; When
$\overline{V_1^{\mathcal{D}}V_2^{\mathcal{D}}}\geq\overline{V_1^{\mathcal{D}}V_3^{\mathcal{D}}}$ and $M_1^{\mathcal{D}},M_2^{\mathcal{D}}\in\mathcal{P}^{\mathcal{D}}$,  we have $\phi_1=\arcsin\left(\frac{\overline{V_1^{\mathcal{D}}M_{13}^{\mathcal{D}}}}{\overline{V_1^{\mathcal{D}}M_1^{\mathcal{D}}}}\right)$,
$\phi_2=\arcsin\left(\frac{\overline{V_1^{\mathcal{D}}M_{12}^{\mathcal{D}}}}{\overline{V_1^{\mathcal{D}}M_1^{\mathcal{D}}}}\right)$.

Regarding the probability $P\left(\mathcal{E}|\sqrt{E_{\mathcal{R}}}x_{\mathcal{R}}=b,T_2\right)$, we have
\begin{equation}\label{eq:perfect2}
P\left(\mathcal{E}|\sqrt{E_{\mathcal{R}}}x_{\mathcal{R}}=b,T_2\right)=1-P_{w_3}\left(d_{21}^{\mathcal{D}},d_{22}^{\mathcal{D}},\phi_1,\phi_2,\phi_3,\phi_4\right),
\end{equation}
where $\phi_3=\angle M_1^{\mathcal{D}}V_2^{\mathcal{D}}M_{2}^{\mathcal{D}}$ and
$\phi_4=\angle V_2^{\mathcal{D}}M_1^{\mathcal{D}}M_2^{\mathcal{D}}$. Further, the determinations of $\phi_1,\phi_2$ in \eqref{eq:perfect2} are as follows. When $\overline{V_1^{\mathcal{D}}V_2^{\mathcal{D}}}<\overline{V_1^{\mathcal{D}}V_3^{\mathcal{D}}}$ and $M_1^{\mathcal{D}},M_2^{\mathcal{D}}\in\mathcal{P}^{\mathcal{D}}$, we have $\phi_1=\arcsin\left(\frac{\overline{V_2^{\mathcal{D}}M_{12}^{\mathcal{D}}}}{\overline{V_2^{\mathcal{D}}M_1^{\mathcal{D}}}}\right)$ and $\phi_2=\pi-\arcsin\left(\frac{\overline{V_2^{\mathcal{D}}M_{24}^{\mathcal{D}}}}{\overline{V_2^{\mathcal{D}}M_2^{\mathcal{D}}}}\right)$. Otherwise, we have $\phi_1=\arcsin\left(\frac{\overline{V_2^{\mathcal{D}}M_{12}^{\mathcal{D}}}}{\overline{V_2^{\mathcal{D}}M_1^{\mathcal{D}}}}\right)$ and
$\phi_2=\arcsin\left(\frac{\overline{V_2^{\mathcal{D}}M_{24}^{\mathcal{D}}}}{\overline{V_2^{\mathcal{D}}M_2^{\mathcal{D}}}}\right)$.

When the relay detects the received signal unsuccessfully, the reference points in Eq.~\eqref{eq:correct_RP_dest} will change according to the incorrect relay decisions. In particular, if sources transmit $T_1$ and the relay wrongly forwards $b,-b,a$, then the reference point $V_1^{\mathcal{D}}$ in Fig.~\ref{figure_dest} will change to $V_5^{\mathcal{D}}$, $V_6^{\mathcal{D}}$, and $V_7^{\mathcal{D}}$, respectively; If sources transmit $T_2$ and the relay wrongly forwards $a,-b,-a$, then the reference point $V_2^{\mathcal{D}}$ in Fig.~\ref{figure_dest} will change to $V_8^{\mathcal{D}}$, $V_9^{\mathcal{D}}$, and $V_{10}^{\mathcal{D}}$, respectively.

In the case when $T_1$ is transmitted by the sources, and the relay wrongly forwards $b,-b,a$, the error probabilities that destination makes wrong decisions are given by
\begin{equation}\label{eq:nonperfect1}
\begin{aligned}
P\left(\mathcal{E}|\sqrt{E_{\mathcal{R}}}x_{\mathcal{R}}=k_1,T_1\right)=\left\{
\begin{array}{c}
  1-P_{w_4}(d_{j1}^{\mathcal{D}},\phi_1,\phi_2),\hspace{2mm}\text{when $V_j^{\mathcal{D}} \in \Omega_{V_1^{\mathcal{D}}}$}, \\
  1-P_{w_1}(d_{j1}^{\mathcal{D}},\phi_1,\phi_2),\hspace{2mm}\text{when $V_j^{\mathcal{D}} \not\in \Omega_{V_1^{\mathcal{D}}}$},
\end{array}
\right.
\end{aligned}
\end{equation}
where when $k_1=b,-b,a$, then $j=5,6,7$, respectively.

In the case when $T_2$ is transmitted by the sources, and the relay wrongly forwards $a,-b,-a$, the error probabilities that destination makes wrong decisions are given by
\begin{equation}\label{eq:nonperfect2}
\begin{aligned}
P\left(\mathcal{E}|\sqrt{E_{\mathcal{R}}}x_{\mathcal{R}}=k_2,T_2\right)=\left\{
\begin{array}{c}
  1-P_{w_3}(d_{l1}^{\mathcal{D}},d_{l2}^{\mathcal{D}},\phi_1,\phi_2,\phi_3,\phi_4,1,1),\hspace{2mm}\text{when $V_l^{\mathcal{D}}\in\Omega_{V_2^{\mathcal{D}}}$}, \\
  1-P_{w_3}(d_{l2}^{\mathcal{D}},d_{l1}^{\mathcal{D}},\phi_1,\phi_2,\phi_3,\phi_4,1,1),\hspace{2mm}\text{when $V_l^{\mathcal{D}}\not\in\Omega_{V_2^{\mathcal{D}}}$},
\end{array}
\right.
\end{aligned}
\end{equation}
where when $k_1=a,-b,-a$, then $l=8,9,10$, respectively. Note that the determinations on the related angles $\phi_1,\phi_2,\phi_3$, and $\phi_4$
are shown in Table I and Table II.
%======================================================
\begin{table}[!h]\label{table:non1}
\tiny
  \centering
  \caption{Correspondence Parameters of $P\left(\mathcal{E}|\sqrt{E_{\mathcal{R}}}x_{\mathcal{R}}=k_1,T_1\right)$}
  \hspace{-1.81989cm}
    \begin{tabular}{|c|c|c|c|c|c|}
    \hline
    \multicolumn{ 3}{|c|}{$V_j^{\mathcal{D}}\in\Omega_{V_1^{\mathcal{D}}}$} & \multicolumn{ 3}{|c|}{$V_j^{\mathcal{D}}\not\in\Omega_{V_1^{\mathcal{D}}}$} \\
    \hline
    \multicolumn{ 2}{|c|}{$M_1,M_2\not\in\mathcal{P}^\mathcal{D}$} & $M_1,M_2\in\mathcal{P}^\mathcal{D}$     & \multicolumn{ 2}{|c|}{$M_1,M_2\not\in\mathcal{P}^\mathcal{D}$} & $M_1,M_2\in\mathcal{P}^\mathcal{D}$ \\
    \hline
    $\overline{V_1^{\mathcal{D}}V_2^{\mathcal{D}}}<\overline{V_1^{\mathcal{D}}V_3^{\mathcal{D}}}$     & $\overline{V_1^{\mathcal{D}}V_2^{\mathcal{D}}}\geq\overline{V_1^{\mathcal{D}}V_3^{\mathcal{D}}}$     & -     & $\overline{V_1^{\mathcal{D}}V_2^{\mathcal{D}}}<\overline{V_1^{\mathcal{D}}V_3^{\mathcal{D}}}$    & $\overline{V_1^{\mathcal{D}}V_2^{\mathcal{D}}}\geq\overline{V_1^{\mathcal{D}}V_3^{\mathcal{D}}}$    & - \\
    \hline
    \tabincell{c}{$\phi_1=\angle V_j^{\mathcal{D}}M_1^{\mathcal{D}}M_{12}^{\mathcal{D}}$\\$\phi_2=\pi-\angle V_j^{\mathcal{D}}M_1^{\mathcal{D}}M_{13}^{\mathcal{D}}$}&
    & \tabincell{c}{$\phi_1=\pi-\angle V_j^{\mathcal{D}}M_1^{\mathcal{D}}M_{12}^{\mathcal{D}}$\\$\phi_2=\angle V_j^{\mathcal{D}}M_1^{\mathcal{D}}M_{13}^{\mathcal{D}}$}&
    & \tabincell{c}{$\phi_1=\angle V_j^{\mathcal{D}}M_1^{\mathcal{D}}M_{12}^{\mathcal{D}}$\\$\phi_2=\angle V_j^{\mathcal{D}}M_1^{\mathcal{D}}M_{13}^{\mathcal{D}}$}&
    & \tabincell{c}{$\phi_1=\pi-\angle V_j^{\mathcal{D}}M_1^{\mathcal{D}}M_{12}^{\mathcal{D}}$\\$\phi_2=\angle V_j^{\mathcal{D}}M_1^{\mathcal{D}}M_{13}^{\mathcal{D}}$}&
    & \tabincell{c}{$\phi_1=\angle V_j^{\mathcal{D}}M_1^{\mathcal{D}}M_{13}^{\mathcal{D}}$\\$\phi_2=\pi-\angle V_j^{\mathcal{D}}M_1^{\mathcal{D}}M_{12}^{\mathcal{D}}$}&
    & \tabincell{c}{$\phi_1=\angle V_j^{\mathcal{D}}M_1^{\mathcal{D}}M_{12}^{\mathcal{D}}$\\$\phi_2=\pi-\angle V_j^{\mathcal{D}}M_1^{\mathcal{D}}M_{13}^{\mathcal{D}}$}& \\
    \hline
    \end{tabular}
    \vspace{-8ex}
  \label{tab:addlabel}%
\end{table}%
\begin{table}[!h]\label{table:non2}
  \centering
  \tiny
  \caption{Correspondence Parameters of $P\left(\mathcal{E}|\sqrt{E_{\mathcal{R}}}x_{\mathcal{R}}=k_2,T_2\right)$}
    \begin{tabular}{|c|c|c|c|c|}
    \hline
    $V_l^{\mathcal{D}}\in\Omega_{V_2^{\mathcal{D}}}$     & \multicolumn{4}{|c|}{$V_l^{\mathcal{D}}\not\in\Omega_{V_2^{\mathcal{D}}}$}        \\
    \hline
    -     & \multicolumn{2}{|c|}{$V_l^{\mathcal{D}}$ is above line $\overline{M_1^{\mathcal{D}}M_2^{\mathcal{D}}}$} & \multicolumn{2}{|c|}{$V_l^{\mathcal{D}}$ is below line $\overline{M_1^{\mathcal{D}}M_2^{\mathcal{D}}}$} \\
    \hline
    -     & $M_1^{\mathcal{D}},M_2^{\mathcal{D}}\not\in\mathcal{P}^{\mathcal{D}}$     & $M_1^{\mathcal{D}},M_2^{\mathcal{D}}\in\mathcal{P}^{\mathcal{D}}$     & $M_1^{\mathcal{D}},M_2^{\mathcal{D}}\not\in\mathcal{P}^{\mathcal{D}}$     & $M_1^{\mathcal{D}},M_2^{\mathcal{D}}\in\mathcal{P}^{\mathcal{D}}$ \\
    \hline
    \tabincell{c}{$\phi_1=\angle V_l^{\mathcal{D}}M_1^{\mathcal{D}}M_{12}^{\mathcal{D}}$\\$\phi_2=\angle V_l^{\mathcal{D}}M_2^{\mathcal{D}}M_{24}^{\mathcal{D}}$\\$\phi_3=\angle
V_l^{\mathcal{D}}M_1^{\mathcal{D}}M_2^{\mathcal{D}}$\\$\phi_4=\angle M_l^{\mathcal{D}}V_1^{\mathcal{D}}M_2^{\mathcal{D}}$}&
    & \tabincell{c}{$\phi_1=\angle V_l^{\mathcal{D}}M_1^{\mathcal{D}}M_2^{\mathcal{D}}$\\$\phi_2=\pi-\angle
V_l^{\mathcal{D}}M_1^{\mathcal{D}}M_{12}^{\mathcal{D}}$\\$\phi_3=\pi-\angle V_l^{\mathcal{D}}M_2^{\mathcal{D}}M_1^{\mathcal{D}}$\\$\phi_4=\pi-\angle
V_l^{\mathcal{D}}M_2^{\mathcal{D}}M_{24}^{\mathcal{D}}$}&
    & \tabincell{c}{$\phi_1=\pi-\angle V_l^{\mathcal{D}}M_2^{\mathcal{D}}M_1^{\mathcal{D}}$\\$\phi_2=\pi-\angle
V_l^{\mathcal{D}}M_2^{\mathcal{D}}M_{24}^{\mathcal{D}}$\\$\phi_3=\angle V_l^{\mathcal{D}}M_1^{\mathcal{D}}M_2^{\mathcal{D}}$\\$\phi_4=\pi-\angle
V_l^{\mathcal{D}}M_1^{\mathcal{D}}M_{12}^{\mathcal{D}}$}&
    & \tabincell{c}{$\phi_1=\pi-\angle V_l^{\mathcal{D}}M_2^{\mathcal{D}}M_{24}^{\mathcal{D}}$\\$\phi_2=\angle
V_l^{\mathcal{D}}M_2^{\mathcal{D}}M_1^{\mathcal{D}}$\\$\phi_3=\pi-\angle V_l^{\mathcal{D}}M_1^{\mathcal{D}}M_{12}^{\mathcal{D}}$\\$\phi_4=\pi-\angle
V_l^{\mathcal{D}}M_1^{\mathcal{D}}M_2^{\mathcal{D}}$}&
    & \tabincell{c}{$\phi_1=\pi-\angle V_l^{\mathcal{D}}M_1^{\mathcal{D}}M_{12}^{\mathcal{D}}$\\$\phi_2=\pi-\angle
V_l^{\mathcal{D}}M_1^{\mathcal{D}}M_2^{\mathcal{D}}$\\$\phi_3=\angle V_l^{\mathcal{D}}M_2^{\mathcal{D}}M_{24}^{\mathcal{D}}$\\$\phi_4=\angle
V_l^{\mathcal{D}}M_2^{\mathcal{D}}M_1^{\mathcal{D}}$}& \\
    \hline
    \end{tabular}%
    \vspace{-5ex}
  \label{tab:addlabel}%
\end{table}%
%===========================================
\subsection{SPER Based on Coordinate Transformation}
We observe that the SPER based on wedge probabilities is diversified into several cases due to varying channel coefficients. Although the SPER result is accurate by using the wedge probability method, the calculations of the SPER could be very complicated. Here, we will propose a coordinate transformation method to derive the SPER, which reduces the calculation complexity by sacrificing a little accuracy in the low SNR region. Specifically, our coordinate transformation method transforms the original parallelogram geometry in the wedge potability method to a rectangle geometry, based on which, we determine the decision regions of the new constellations. The following lemma derives the coordinate transformation matrix at relay node.
\begin{lemma}\label{lemma:relay}
The coordinate transformation matrix $\mathbf{C}$, which transforms the exact parallelogram-shaped IRC to a rectangle centered at origin point, and preserve the length of each side in exact IRC is given by
\begin{equation}\label{eq:C_relay}
\mathbf{C}=\mathbf{Q}\mathbf{A}^{-1},
\end{equation}
where $\mathbf{Q}$ is the eigenvector matrix for $\mathbf{B}=\mathbf{A}^T\Sigma^{-1}\mathbf{A}$ given in \eqref{eq:B_relay}, shown as
\begin{equation}\label{eq:Q_relay}
\mathbf{Q}=\left[
 \begin{array}{cc}
\frac{\mathbf{B}(1,2)}{(\lambda_1-\lambda_2)\sqrt{\frac{\mathbf{B}(1,1)-\lambda_1}{\lambda_2-\lambda_1}}} & \sqrt{\frac{\mathbf{B}(1,1)-\lambda_1}{\lambda_2-\lambda_1}} \\
\sqrt{\frac{\mathbf{B}(1,1)-\lambda_1}{\lambda_2-\lambda_1}} & -\frac{\mathbf{B}(1,2)}{(\lambda_1-\lambda_2)\sqrt{\frac{\mathbf{B}(1,1)-\lambda_1}{\lambda_2-\lambda_1}}} \\
\end{array}
\right],\hspace{1mm}\text{and}\hspace{1mm}
\mathbf{A}^{-1}=\left[\begin{array}{cc}
\frac{\Re\{h_{1\mathcal{R}}\}}{2|h_{1\mathcal{R}}|} & \frac{\Re\{h_{2\mathcal{R}}\}}{2|h_{2\mathcal{R}}|} \\\frac{\Im\{h_{1\mathcal{R}}\}}{2|h_{1\mathcal{R}}|} & \frac{\Im\{h_{2\mathcal{R}}\}}{2|h_{2\mathcal{R}}|} \\\end{array}\right].
\end{equation}
where the eigenvalues $\lambda_1$ and $\lambda_2$ are derived in~\eqref{eq:eigenvalue}.

\end{lemma}

\emph{Proof:} The proof of Lemma~\ref{lemma:relay} is illustrated in Appendix \ref{appendix_relay}.\hfill$\blacksquare$

\begin{figure}
\centering
\includegraphics[width=3.5in]{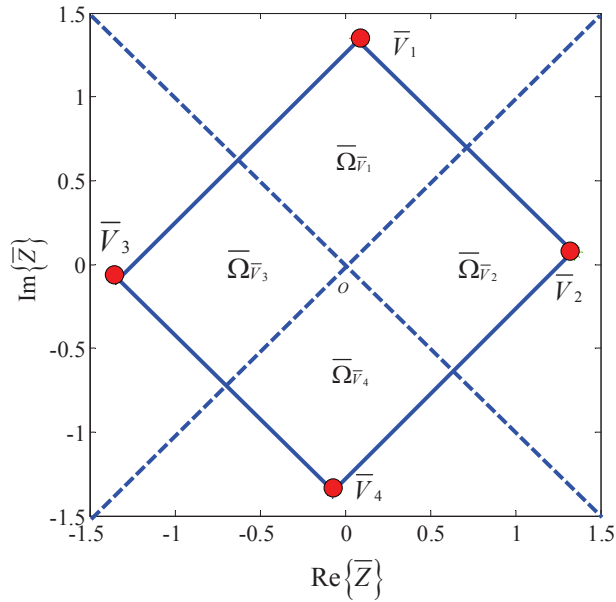}
\caption{Received Constellation at Relay after Coordinate Transformation.}\label{fig:trans_relay}
\vspace{-5ex}
\end{figure}

Now, we will determine the decision regions of the new IRC. Denote $\overline{\mathbf{V}}_i$ the constellation point and $\bar{Z}$ the received signal point $(\Re\{y_{\mathcal{R}}\},\Im\{y_{\mathcal{R}}\})$ after coordinate transformation by matrix $\mathbf{B}$, respectively. Denote $\bar{\Omega}_{\bar{\mathbf{V}}_i}$ the decision region corresponding to $\overline{\mathbf{V}}_i$. The boundary is defined based on Voronoi rule
\begin{equation}\label{eq:decision}
\begin{aligned}
&\bar{\Omega}_{\bar{\mathbf{V}}_1}:\Re\{\bar{Z}\}-\Im\{\bar{Z}\}>0 \bigcap \Re\{\bar{Z}\}+\Im\{\bar{Z}\}\leq 0,
\hspace{2mm}\bar{\Omega}_{\bar{\mathbf{V}}_2}:\Re\{\bar{Z}\}-\Im\{\bar{Z}\}>0 \bigcap \Re\{\bar{Z}\}+\Im\{\bar{Z}\}>0,\\
&\bar{\Omega}_{\bar{\mathbf{V}}_3}:\Re\{\bar{Z}\}-\Im\{\bar{Z}\}\leq 0 \bigcap \Re\{\bar{Z}\}+\Im\{\bar{Z}\}> 0,
\hspace{2mm}\bar{\Omega}_{\bar{\mathbf{V}}_4}:\Re\{\bar{Z}\}-\Im\{\bar{Z}\}\leq 0 \bigcap \Re\{\bar{Z}\}+\Im\{\bar{Z}\}\leq 0.
\end{aligned}
\end{equation}
In Fig.~\ref{fig:trans_relay}, we present the constellation after coordinate transformation and its decision regions.

From~\eqref{eq:decision}, we can see that the decision regions are regular geometry with simple decision boundary lines\footnote{Note that the real decision boundary lines are slightly different from the perpendicular bisectors of Voronoi diagram after coordinate transformation. This is why the SPER results of coordinate transformation have notably little difference in low SNR comparing with its counterpart of exact constellation. As the SNR goes larger, the real decision boundary lines are coincide with the perpendicular bisectors.}. Thus, it will greatly reduce the computational complexity. The probabilities that relay detects the received signal when $T_1$ is sent are shown as
\begin{equation}\label{eq:relay1}
\begin{aligned}
P(\sqrt{E_{\mathcal{R}}}x_{\mathcal{R}}=a|T_1)&=\int_{0}^{\infty}d(\Im\{\bar{Z}\})\int_{-\Im\{\bar{Z}\}}^{\Im\{\bar{Z}\}}
f_{\overline{Z}}\left(\bar{Z};\Re\left\{\overline{\mathbf{V}}_i\right\},\Im\left\{\overline{\mathbf{V}}_i\right\}\right)d(\Re\{\bar{Z}\}),\\
\end{aligned}
\end{equation}
where $f_{\overline{Z}}(\cdot)$ is the probability density function of $\overline{Z}$ given in~\eqref{eq:pdf2}, $\overline{\mathbf{V}}_i(1)$ and $\overline{\mathbf{V}}_i(2)$ represent the horizontal coordinate and vertical coordinate of $\overline{\mathbf{V}}_i$, respectively. Similar to \eqref{eq:relay1}, we can obtain $P(\sqrt{E_{\mathcal{R}}}x_{\mathcal{R}}=k|T_1)$ for $k \in \{\pm b, -a\}$ and the probabilities that relay detects the received signal when $T_2$ is sent.

Similar to the error probability analysis at relay, we will show the results of coordinate transformation and the error probability based on the new constellation at destination. In the following lemma, we present the coordinate transformation matrix $\mathbf{C}_{\mathcal{D}}$ at destination.

\begin{lemma}\label{lemma:dest}
At destination, the coordinate transformation matrix $\mathbf{C}_{\mathcal{D}}$, which transform the exact IDC to a rectangle centered at origin point, and preserve the length of each side in exact parallelogram-shaped IDC is given by
\begin{equation}\label{eq:C_relay}
\mathbf{C}_{\mathcal{D}}=\mathbf{Q}_{\mathcal{D}}\mathbf{A}^{-1}_{\mathcal{D}},
\end{equation}
where $\mathbf{Q}_{\mathcal{D}}$ is the eigenvector matrix for $\mathbf{B}_{\mathcal{D}}$ shown as
\begin{equation}\label{eq:B_dest}
\mathbf{B}_{\mathcal{D}}=\frac{2}{\beta_{\mathcal{D}}^2\sigma^2}\left[\begin{array}{cc}d_1^2(a+b)^2|h_{\mathcal{RD}}|^2+4d_1^2|h_{2\mathcal{D}}|^2 & d_1d_2(b^2-a^2)|h_{\mathcal{RD}}|^2-4d_1d_2|h_{1\mathcal{D}}||h_{2\mathcal{D}}| \\d_1d_2(b^2-a^2)|h_{\mathcal{RD}}|^2-4d_1d_2|h_{1\mathcal{D}}||h_{2\mathcal{D}}| & d_2^2(b-a)^2|h_{\mathcal{RD}}|^2+4d_2^2|h_{2\mathcal{D}}|^2 \\\end{array} \right],
\end{equation}
where $\beta_{\mathcal{D}}=h_{\mathcal{RD}}\left(|h_{1\mathcal{D}}|(a+b)+|h_{2\mathcal{D}}|(b-a)\right)$, $d_1=\sqrt{4|h_{1\mathcal{D}}|^2+(a-b)^2|h_{\mathcal{RD}}|^2}$ and $d_2=\sqrt{4|h_{2\mathcal{D}}|^2+(a+b)^2|h_{\mathcal{RD}}|^2}$. And
\begin{equation}\label{eq:Q_relay}
\mathbf{Q}_{\mathcal{D}}=\left[
\begin{array}{cc}
\frac{\mathbf{B}_{\mathcal{D}}(1,2)}{(\lambda_1^{\mathcal{D}}-\lambda_2^{\mathcal{D}})\sqrt{\frac{\mathbf{B}_{\mathcal{D}}(1,1)-\lambda_1^{\mathcal{D}}}{\lambda_2^{\mathcal{D}}-\lambda_1^{\mathcal{D}}}}} & \sqrt{\frac{\mathbf{B}_{\mathcal{D}}(1,1)-\lambda_1^{\mathcal{D}}}{\lambda_2^{\mathcal{D}}-\lambda_1^{\mathcal{D}}}} \\
\sqrt{\frac{\mathbf{B}_{\mathcal{D}}(1,1)-\lambda_1^{\mathcal{D}}}{\lambda_2^{\mathcal{D}}-\lambda_1^{\mathcal{D}}}} & -\frac{\mathbf{B}_{\mathcal{D}}(1,2)}{(\lambda_1^{\mathcal{D}}-\lambda_2^{\mathcal{D}})\sqrt{\frac{\mathbf{B}_{\mathcal{D}}(1,1)-\lambda_1^{\mathcal{D}}}{\lambda_2^{\mathcal{D}}-\lambda_1^{\mathcal{D}}}}} \\
\end{array}
\right],\hspace{1mm}\text{and}\hspace{1mm}
\mathbf{A}^{-1}_{\mathcal{D}}=\left[\begin{array}{cc}(a+b)|h_{\mathcal{RD}}| & -|h_{2\mathcal{D}}| \\(a-b)|h_{\mathcal{RD}}| & -|h_{1\mathcal{D}}| \\\end{array}\right].
\end{equation}
where the eigenvalues $\lambda_1^{\mathcal{D}}$ and $\lambda_2^{\mathcal{D}}$ are eigenvalues of $\mathbf{B}_{\mathcal{D}}$.
\end{lemma}

Lemma \ref{lemma:dest} can be obtained in the same way as we formulated at relay. Now, we will determine the decision regions with the new IDC. Denote $\overline{\mathbf{V}}_i^{\mathcal{D}}$ the constellation point and $\bar{Z}_{\mathcal{D}}$ the received signal after coordinate transformation by matrix $\mathbf{C}_{\mathcal{D}}$. Denote $\bar{\Omega}_{\overline{\mathbf{V}}_i^{\mathcal{D}}}^{\mathcal{D}}$ the decision region corresponding to $\overline{\mathbf{V}}_i^{\mathcal{D}}$. The boundary is defined based on Voronoi rule
\begin{equation}
\small
\begin{aligned}
&\bar{\Omega}_{\overline{\mathbf{V}}_1^{\mathcal{D}}}^{\mathcal{D}}:\bar{Z}_{\mathcal{D}}(1)-\bar{Z}_{\mathcal{D}}(2)>0 \bigcap \bar{Z}_{\mathcal{D}}(1)+\bar{Z}_{\mathcal{D}}(2)\leq 0,\hspace{2mm}\bar{\Omega}_{\overline{\mathbf{V}}_2^{\mathcal{D}}}^{\mathcal{D}}:\bar{Z}_{\mathcal{D}}(1)-\bar{Z}_{\mathcal{D}}(2)>0 \bigcap \bar{Z}_{\mathcal{D}}(1)+\bar{Z}_{\mathcal{D}}(2)>0,\\
&\bar{\Omega}_{\overline{\mathbf{V}}_3^{\mathcal{D}}}^{\mathcal{D}}:\bar{Z}_{\mathcal{D}}(1)-\bar{Z}_{\mathcal{D}}(2)\leq 0 \bigcap \bar{Z}_{\mathcal{D}}(1)+\bar{Z}_{\mathcal{D}}(2)> 0,\hspace{2mm}\bar{\Omega}_{\overline{\mathbf{V}}_4^{\mathcal{D}}}^{\mathcal{D}}:\bar{Z}_{\mathcal{D}}(1)-\bar{Z}_{\mathcal{D}}(2)\leq 0 \bigcap \bar{Z}_{\mathcal{D}}(1)+\bar{Z}_{\mathcal{D}}(2)\leq 0,
\end{aligned}
\end{equation}
where $\bar{Z}_{\mathcal{D}}(1)$ and $\bar{Z}_{\mathcal{D}}(2)$ represent the horizontal coordinate and
vertical coordinate of $\bar{Z}_{\mathcal{D}}$, respectively.

Based on the decision regions of new IDC, the probability that destination detects the received signal successfully and unsuccessfully when $T_1$ is sent are shown respectively as
\begin{equation}\label{eq:dest1}
\begin{aligned}
P\left(\mathcal{E}|\sqrt{E_{\mathcal{R}}}x_{\mathcal{R}}=a,T_1\right)&
=\int_{0}^{\infty}d(\bar{Z}_{\mathcal{D}}(2))\int_{-\bar{Z}_{\mathcal{D}}(2)}^{\bar{Z}_{\mathcal{D}}(2)}
f_{\overline{Z}_{\mathcal{D}}}\left(\bar{Z}_{\mathcal{D}};\overline{\mathbf{V}}_i^{\mathcal{D}}(1),\overline{\mathbf{V}}_i^{\mathcal{D}}(2)\right)d(\bar{Z}_{\mathcal{D}}(1)),\\
\end{aligned}
\end{equation}
where $f_{\overline{Z}_{\mathcal{D}}}(\cdot)$ is the probability density function of $\bar{Z}_{\mathcal{D}}$,
$\overline{\mathbf{V}}_i^{\mathcal{D}}(1)$ and $\overline{\mathbf{V}}_i^{\mathcal{D}}(2)$ represent the horizontal coordinate and vertical coordinate
of $\overline{\mathbf{V}}_i^{\mathcal{D}}$, respectively. Similar to \eqref{eq:dest1}, we can obtain $P\left(\mathcal{E}|\sqrt{E_{\mathcal{R}}}x_{\mathcal{R}}=k,T_1\right)$ for $k\in\{\pm b, -a\}$ and the probabilities that relay detects the received signal when $T_2$ is sent.

Based on \eqref{eq:relay1} and \eqref{eq:dest1}, we can calculate the SPER shown in \eqref{eq:general} based on the constellation after coordinate transformation.

%====================================================Section 5==============================
\section{System Optimization}\label{sec:optimal}
In this section, we first develop a practical method at the relay side to address the error propagation problem. With the designed method, the system is proved to achieve full diversity when the relay implements non-perfect detection. Specifically, we propose a power scaling scheme where the relay power is adaptive to the channel conditions. For such link adaptive relaying (LAR) scheme, we model the complicated MARC system as a degraded virtual one-source one-relay one destination model (triangle model), and show that the relay power should be such to balance the SNRs of source-relay channel and relay-destination channel. Moreover, we formulate a sub-optimal Max-min method to obtain the optimized system parameters $a$ and $b$ that minimize the end-to-end SER.
%=============================================================
\subsection{The Design of Power Scaling Factor at Relay}

Before introducing the design of power scaling factor at relay, we would present the diversity performance of the proposed PANC scheme and the CXNC scheme in the following Proposition.
\begin{theorem}\label{theorem:noalpha}
Without the design of power scaling factor at relay, both the PANC scheme and the CXNC scheme can only achieve one order diversity in the MARC system.
\end{theorem}

\emph{Proof:} Please refer to Appendix \ref{sub:proposition}.

From Theorem \ref{theorem:noalpha}, we can see that the error propagation from the early source-relay hop degrades the performance of the system. In this case, we adopt a power scaling factor at relay to leverage the effect of error propagation by adjusting the relay transmission power according to the channel conditions. Such link adaptive ratio (LAR) was first introduced for the single-source DF system in~\cite{Wang:TCOM08}. However, LAR cannot be directly applied to the multi-user power adapted network coding system.

\begin{figure}\label{fig:virtual}
\centering
\includegraphics[width=4.5in]{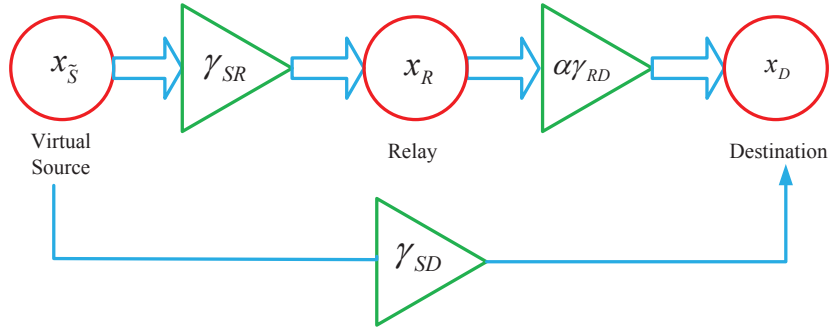}
{\caption{Virtual channel model.}}
\vspace{-5ex}
\end{figure}

To extend the spirit of LAR, we first develop a virtual channel model for the source-relay-destination link, as shown in Fig.~4. In the first phase, the two sources transmit to relay simultaneously. For such multiple-access channel, the upper bound for the instantaneous SER is given by
\begin{equation}
\begin{aligned}
P_{\text{MAC}}\leq P_{\text{MAC}}^U &\triangleq Q_1\left(\sqrt{2E_1|h_{1\mathcal{R}}|^2/\sigma^2}\right)+Q_1\left(\sqrt{2E_2|h_{2\mathcal{R}}|^2/\sigma^2}\right)\\
&+Q_1\left(\sqrt{2|\sqrt{E_1}h_{1\mathcal{R}}+\sqrt{E_2}h_{2\mathcal{R}}|^2/\sigma^2}\right),\\
\end{aligned}
\end{equation}
which can be further approximated as
\begin{equation}
P_{\text{MAC}}^U\approx Q_1\left(\sqrt{2\min\left[E_1|h_{1\mathcal{R}}|^2/\sigma^2, E_2|h_{2\mathcal{R}}|^2/\sigma^2, |\sqrt{E_1}h_{1\mathcal{R}}+\sqrt{E_2}h_{2\mathcal{R}}|^2/\sigma^2\right]}\right),
\end{equation}
which is quite tight when $E_1|h_{1\mathcal{R}}|^2/\sigma^2$,  $E_2|h_{2\mathcal{R}}|^2/\sigma^2$, $|\sqrt{E_1}h_{1\mathcal{R}}+\sqrt{E_2}h_{2\mathcal{R}}|^2/\sigma^2$ and their difference are reasonably large, as the one-dimensional Q-function $Q_1(x)$ decays fast as $x$ grows. The advantage of such approximation is that we can now model the multiple access source-relay channel as a single-input single-output channel with the input being the virtual source message $x_{\tilde{S}}=x_1\boxplus x_2$ and the instantaneous channel SNR being $\gamma_{\mathcal{SR}}\triangleq\min(E_1|h_{1\mathcal{R}}|^2, E_2|h_{2\mathcal{R}}|^2, |\sqrt{E_1}h_{1\mathcal{R}}+\sqrt{E_2}h_{2\mathcal{R}}|^2)$ that is the SNR of the worse source-relay channel\footnote{Note that our approximation is different from the one shown in \cite{Guan:TWC12}, in which the authors consider an orthogonal MARC system.}. The idea of regarding virtual source message as network coded sources' signals is that we implement network coding at relay node. Thus, the virtual transmitting information from source to destination via the aid of relay becomes the same. Likely, we can model the multiple access source-destination channel as a point-to-point channel with the channel SNR as $\gamma_{\mathcal{SD}}$. So far, we have successfully degrade the complex MARC system to a traditional triangle model. Based on the conclusion in~\cite{Wang:TCOM08}, the power scaling factor $\alpha$ with instantaneous $\gamma_{\mathcal{SR}}$ and $\gamma_{\mathcal{RD}}$ is given by
\begin{equation}\label{eq:alpha}
\alpha=\min\left(\frac{\gamma_{\mathcal{SR}}}{\gamma_{\mathcal{RD}}},1\right).
\end{equation}
Note that, instantaneous channel SNR $\gamma_{\mathcal{RD}}$ can be replaced by statistical channel SNR $\bar{\gamma}_{\mathcal{RD}}$. The advantage of using $\bar{\gamma}_{\mathcal{RD}}$ to obtain $\alpha$ is that the relay node does not need the feedback of relay-destination channel. Later, we will show that both instantaneous and statistical relay-destination channel SNR can achieve full diversity in the proposed PANC scheme.

\begin{theorem}\label{tho:theorem1}
Given instantaneous source-relay channel SNR, and instantaneous (or statistical) relay-destination channel SNR, the power scaled PANC scheme can achieve two order diversity, i.e., the full diversity, in the MARC system, while the power scaled conventional NC scheme can only achieve one order diversity.
\end{theorem}

\emph{Proof:} Please refer to Appendix \ref{sub:Diversity_NCPA}.

%=============================================================
\subsection{The Design of Power Adaption Factors}\label{subsection:inst}
From the derivations of the instantaneous system SPER in Section~\ref{sec:error}, we note that the expressions of the SPER depends on the power adaption levels $a$ and $b$ at the relay. To minimize the SPER requires a smart optimization on $a$ and $b$. However, directly minimizing the SPER is very complicated and leads to no closed forms of $a$ and $b$. Here, we propose a sub-optimal criterion for the instantaneous SPER minimization, i.e., maximizing the minimum Euclidean distances of the instantaneous constellation at the destination. The \emph{Max-min} optimization problem under the power constraint is formulated as
\begin{equation}\label{eq:opt}
\begin{aligned}
\left(a^*,b^*\right)&=\mathop{\text{arg}\max }\limits_{a,b}\mathop{\min }\limits_{k,j = 1,2,3,4;k \ne j}\left\{||V_k^{\mathcal{D}} - V_j^{\mathcal{D}}||^2 \right\}\\
&\text{s. t.}\hspace{2mm}a^2+b^2 \leq 2E_{\mathcal{R}}^{\text{ave}},\hspace{2mm}a,b \in \mathbb{R},
\end{aligned}
\end{equation}
where the lengths of parallelogram's two edges are $\overrightarrow{V_1^{\mathcal{D}}V_2^{\mathcal{D}}}=||V_1^{\mathcal{D}}-V_2^{\mathcal{D}}||^2=4E_1|h_{1\mathcal{D}}|^2+|h_{\mathcal{RD}}|^2(a-b)^2$ and $\overrightarrow{V_1^{\mathcal{D}}V_3^{\mathcal{D}}}=4E_2|h_{2\mathcal{D}}|^2+|h_{\mathcal{RD}}|^2(a+b)^2$, and the lengths of diagonals are $\overrightarrow{V_2^{\mathcal{D}}V_3^{\mathcal{D}}}=\left(-2\sqrt{E_1}|h_{1\mathcal{D}}|+2\sqrt{E_2}|h_{2\mathcal{D}}|\right)^2+4|h_{\mathcal{RD}}|^2b^2$ and $\overrightarrow{V_1^{\mathcal{D}}V_4^{\mathcal{D}}}=\left(2\sqrt{E_1}|h_{1\mathcal{D}}|+2\sqrt{E_2}|h_{2\mathcal{D}}|\right)^2+4|h_{\mathcal{RD}}|^2a^2$. Defining $\mathcal{V}$ as the set of $\left\{\overrightarrow{V_1^{\mathcal{D}}V_2^{\mathcal{D}}},\overrightarrow{V_1^{\mathcal{D}}V_3^{\mathcal{D}}},\overrightarrow{V_2^{\mathcal{D}}V_3^{\mathcal{D}}},
\overrightarrow{V_1^{\mathcal{D}}V_4^{\mathcal{D}}}\right\}$, and introducing a variable $u\triangleq\min\{\mathcal{V}\}$, after some manipulations, the \emph{Max-min} problem in~(\ref{eq:opt}) can be further described as a maximization problem
\begin{equation}\label{eq:final_optimal}
\begin{aligned}
\mathop{\max }\limits_{\left(u^*,a^*,b^*\right)}&u\\
\text{s. t.}\hspace{2mm}&-(4E_1|h_{1\mathcal{D}}|^2+|h_{\mathcal{RD}}|^2(a-b)^2)\leq -u,-(4E_2|h_{2\mathcal{D}}|^2+|h_{\mathcal{RD}}|^2(a+b)^2)\leq -u,\\
&-(c_1+4|h_{\mathcal{RD}}|^2b^2)\leq -u,-(c_2+4|h_{\mathcal{RD}}|^2a^2)\leq -u,a^2+b^2 \leq 2E_{\mathcal{R}}^{\text{ave}},
\end{aligned}
\end{equation}
where $c_1=\left(-2\sqrt{E_1}|h_{1\mathcal{D}}|+2\sqrt{E_2}|h_{2\mathcal{D}}|\right)^2$ and $c_2=\left(2\sqrt{E_1}|h_{1\mathcal{D}}|+2\sqrt{E_2}|h_{2\mathcal{D}}|\right)^2$.
Since the objective function is an affine function and the constraints are quadratic functions of $a$ and $b$ in (\ref{eq:final_optimal}), it is a convex optimization problem. We can adopt the Lagrange Multiplier method to obtain the solutions. The Lagrange equation is given by
\begin{multline}
L(a,b,u,\mu_1,\mu_2,\mu_3,\mu_4,\mu_5)=u+\mu_1(u-4E_1|h_{1\mathcal{D}}|^2-|h_{\mathcal{RD}}|^2(a-b)^2)+\mu_2(u-4E_2|h_{2\mathcal{D}}|^2-|h_{\mathcal{RD}}|^2(a+b)^2)
\\+\mu_3(u-c_1-4|h_{\mathcal{RD}}|^2b^2)+\mu_4(u-c_2-4|h_{\mathcal{RD}}|^2a^2)+\mu_5(a^2+b^2-E_{\mathcal{R}}^{\text{ave}}).
\end{multline}

After some intermediate manipulations, we obtain one of the optimal solutions for the problem in (\ref{eq:final_optimal}) in the case
$\mu_1=\mu_2=0,\mu_3 \neq 0,\mu4 \neq 0$, and $\mu5 \neq 0$ as follows.
\begin{equation}\label{eq:ab1}
(a^*,b^*)=\left(\sqrt{E_{\mathcal{R}}^{\text{ave}}+\frac{c_1-c_2}{8|h_{\mathcal{RD}}|^2}},\sqrt{E_{\mathcal{R}}^{\text{ave}}+\frac{c_2-c_1}{8|h_{\mathcal{RD}}|^2}}\right).
\end{equation}
Note that there are total $32$ optimal solutions by adopting different $\mu_1,\mu_2,\mu_3,\mu_4$ and $\mu_5$, which can be obtained similarly by adopting the KKT conditions. But one of these solutions is sufficient to achieve the optimal $u$.

We can also formulate the \emph{Max-min} method for the coordinate transformed IDC proposed in Subsection B of Section \ref{sec:error}, which simplifies the calculations of the SPER. Here, based on the coordinate transformed IDC, we can also simplify the optimization process of $a$ and $b$. Specifically in the \emph{Max-min} problem described in (\ref{eq:opt}), we only consider the the two edges of the rectangle in the coordinate transformed IDC, since the two edges are always less than the diagonals of the rectangle. Likewise, we can implement the \emph{Max-min} method to obtain the optimal $a$ and $b$ based on the IDC after coordinated transforming, one of the solution is
\begin{equation}\label{eq:ab2}
\begin{aligned}
a^*&=\frac{1}{2}\left(\sqrt{\frac{2\left(E_{\mathcal{R}}^{\text{ave}}|h_{\mathcal{RD}}|^2+E_2|h_{2\mathcal{D}}|^2-E_1|h_{1\mathcal{D}}|^2\right)}
{|h_{\mathcal{RD}}|^2}}+\sqrt{\frac{2\left(E_{\mathcal{R}}^{\text{ave}}|h_{\mathcal{RD}}|^2+E_1|h_{1\mathcal{D}}|^2-E_2|h_{2\mathcal{D}}|^2\right)}
{|h_{\mathcal{RD}}|^2}}\right),\\
b^*&=\frac{1}{2}\left(\sqrt{\frac{2\left(E_{\mathcal{R}}^{\text{ave}}|h_{\mathcal{RD}}|^2+E_1|h_{1\mathcal{D}}|^2-E_2|h_{2\mathcal{D}}|^2\right)}
{|h_{\mathcal{RD}}|^2}}-\sqrt{\frac{2\left(E_{\mathcal{R}}^{\text{ave}}|h_{\mathcal{RD}}|^2+E_2|h_{2\mathcal{D}}|^2-E_1|h_{1\mathcal{D}}|^2\right)}
{|h_{\mathcal{RD}}|^2}}\right).
\end{aligned}
\end{equation}

%=================================================================Section 6=============================================================
\section{Simulation Results}\label{sec:sim}
In this section, we evaluate the performance of the proposed PANC scheme by simulations. Consider a two-dimensional cartesian coordinate system, where nodes $\mathcal{S}_1$, $\mathcal{S}_2$ and $\mathcal{D}$ are located at $(0,\frac{\sqrt{3}}{3})$, $(0,-\frac{\sqrt{3}}{3})$, and $(1,0)$, respectively. The relay node is moving from origin point $(0,0)$ to $(1,0)$ at X-axis. Throughout our simulations, we use the path loss model $\gamma_{ij}=d_{ij}^{-3}$, where $\gamma_{ij}$ is the channel gain and $d_{ij}$ is the distance between two terminals, where $i \in \{\mathcal{S}_1,\mathcal{S}_2,\mathcal{R}\}$ and $j \in \{\mathcal{R},\mathcal{D}\}$. The average SNR range is $[0,30]$ dB. To simplify the expression in the legends of simulation results, 'sim' stands for Monte-Carlo simulation result, 'thy' stands for theoretical derivation result.

\begin{figure}
\begin{center}
\includegraphics[width=4.5in]{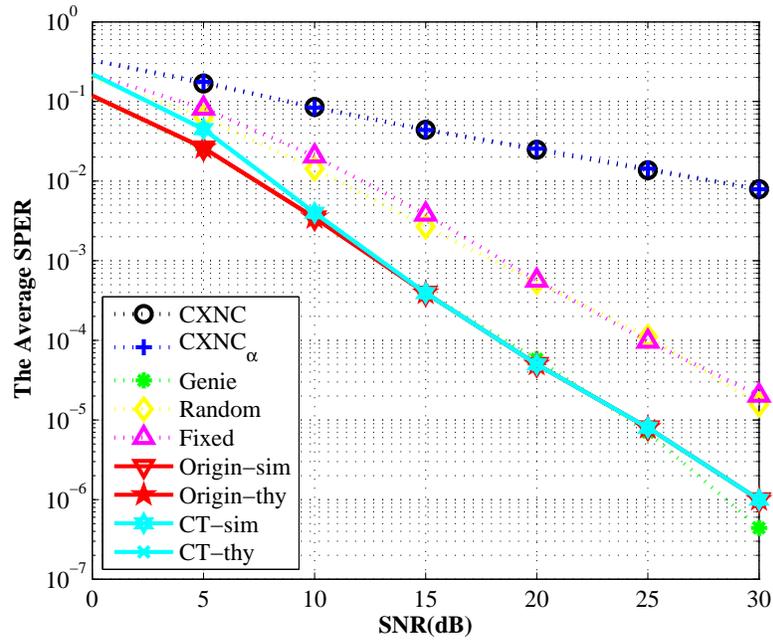}
\vspace{-3ex}
\caption{Error performance with strong source-relay channel (Best viewed in color.).}\label{fig:sim1}
\end{center}
\vspace{-5ex}
\end{figure}

\begin{figure}
\begin{center}
\includegraphics[width=4.5in]{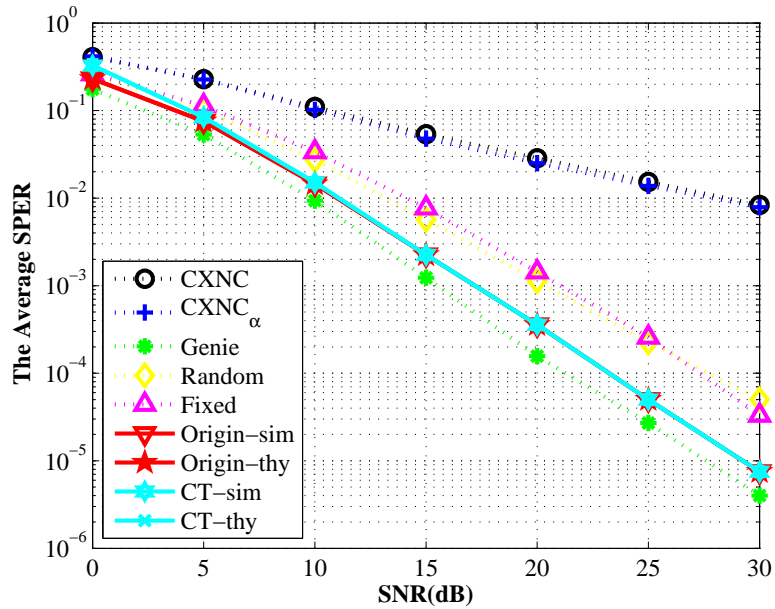}
\vspace{-3ex}
\caption{Error performance in a symmetric network (Best viewed in color.).}\label{fig:sim2}
\vspace{-3ex}
\end{center}
\end{figure}

\begin{figure}
\begin{center}
\includegraphics[width=4.5in]{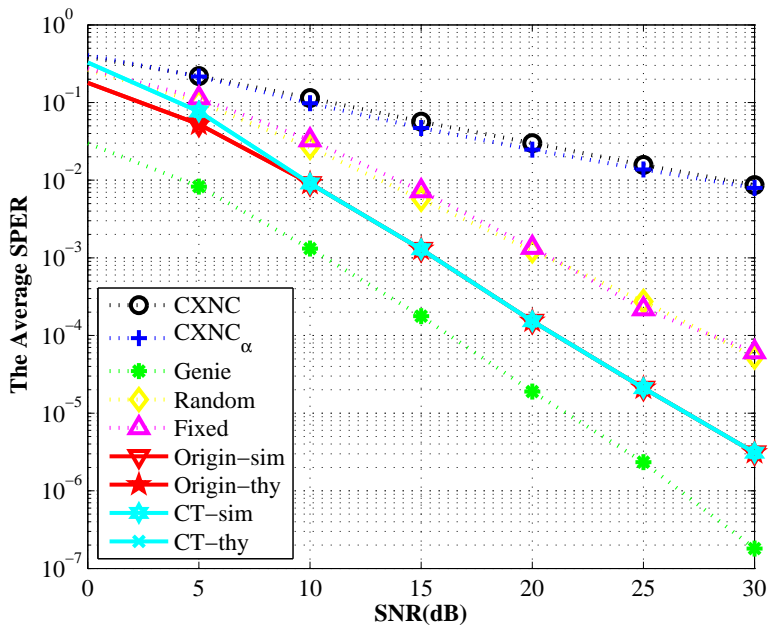}
\caption{Error performance with strong relay-destination channel (Best viewed in color.).}\label{fig:sim3}
\end{center}
\vspace{-5ex}
\end{figure}

In order to investigate the performance of our proposed scheme comprehensively, we consider the relay located at different locations resulting in different channel scenarios. Firstly, we consider the relay is located at $(0,0)$, so the relay is close to the sources, i.e., asymmetric network with strong source-relay channel, shown in Fig.~\ref{fig:sim1}. Then, we consider the relay is located at $(\frac{1}{3},0)$, so the distance between source and relay is equal to the distance between relay and destination, i.e., a symmetric network, shown in Fig.~\ref{fig:sim2}. Finally, we consider the relay is located at $(0.8,0)$, so the distance between sources and relay is larger than the distance between relay to destination, i.e., asymmetric network with strong relay-destination channel, shown in Fig.~\ref{fig:sim3}.

In each realization of nodes locations, we simulate the following schemes to demonstrate the performance of the proposed PANC scheme: (1) the SPER performance of original received constellation with optimized $a$ and $b$ given in \eqref{eq:ab1}, for both monte carlo simulation and theoretically derived result, denoted by \emph{Origin-sim} and \emph{Origin-thy}, respectively. (2) the SPER performance of received constellation after coordinate transformation with optimized $a$ and $b$ given in \eqref{eq:ab2}, for both monte carlo simulation and theoretically derived result, denoted by \emph{CT-sim} and \emph{CT-thy}, respectively.

As the references, we also simulate the following schemes: (1) the SPER performance of CXNC scheme~\cite{Zhang:JSAC09,Popovski:ICC06} in which the relay transmits XORed signal to the destination in the second transmission phase, denoted by \emph{CXNC}; (2) The SPER performance of CXNC scheme in which the relay transmits power scaled XORed signal to the destination in the second transmission phase, denoted by \emph{CXNC$_\alpha$}; (3) The SPER performance of genie-aided PANC scheme which is under the assumption that the symbols are perfectly detected at relay, denoted by \emph{Genie}; (4) The SPER performance of PANC scheme with randomly generated $a$ and $b$, in which $a$ is a uniformly distributed random variable in $[0,\sqrt{2}]$ and $b=\sqrt{2E_{\mathcal{R}}-a^2}$, and both $a$ and $b$ vary in different instantaneous CSI, denoted by \emph{Random}; (5) the SPER performance of PANC scheme with fixed chosen $a$ and $b$, in which $a$ is a uniformly distributed random variable in $[0,\sqrt{2}]$ and $b=\sqrt{2E_{\mathcal{R}}-a^2}$. Once both $a$ and $b$ are generated, they will remain the same for all CSI, denoted by \emph{Fixed}.

Firstly, the CXNC without power scaling can only achieve one order diversity, since the error propagation from the early source-relay hop degrades the performance of the system as Theorem \ref{theorem:noalpha} indicates. The CXNC with power scaling cannot either achieve full diversity due to the multi-user interference of non-orthogonal MARC as Theorem \ref{tho:theorem1} infers. The proposed PANC scheme with the power scaled factor can achieve full diversity no matter what power levels it adopts at relay node, which verify the proof of Theorem \ref{tho:theorem1}. The \emph{Genie} method plays as the benchmark of the system performance, since it assumes that a genie exists at relay and guarantee that the relay transmit correct information to the destination. We can conclude from the simulation results that the proposed PANC scheme with the design of allocating different power levels and adopting the power scaling factor can achieve full diversity in MARC system.

In addition,  the SPER performance based on new coordinate is larger than the SPER based on original coordinate in low SNR (e.g., from $0$dB to $5$dB), and coincides after $10$dB. The reason for such phenomenon is that the detector based on the new coordinate is sub-optimal comparing to the optimal ML detection in~\eqref{eq:detection}. Thus, the error performance based on the new coordinate is poorer than its counterpart based on the original coordinate. As the SNR grows, the minimum Euclidean distance between constellation points in both original and new coordinates increases. So the error performances become perfectly matched. The theoretically derived results of both original coordinate and transformed coordinate match the Monte-Carlo simulation results. It infers that the closed-form SPER expression is accurate by wedge error calculation method and the CT method.

Notice that, allocating different power levels at relay may vary the coding gain of the system. In particular, the SPER performance of optimized $a$ and $b$ based on the Maxmin method has the best coding gain for both original coordinate and CT case. It infers that our Maxmin is performance efficient and it has low complexity compared with exhaustive search. The SPER performance of random or fixed chosen $a$ and $b$ has poorer coding gain performance because they are not adaptive to the instantaneous CSI compared with the optimized $a$ and $b$.

Since there exists error propagation from early source-relay hop to destination, we notice that the gap between the SPER performance of genie-aided PANC scheme and the SPER performance of original received constellation with optimized $a$ and $b$ is different with three relay location realization. In particular, such gap is quite small when we have a strong source-relay channel, and the gap becomes bigger as the relay moves further from source and closer to the destination. The reason is that with strong source-relay channel, the relay will generates more reliable information and thus degrades the influence of error propagation to destination.

%=================================================================Section 7=============================================================
\section{Conclusion }
In this paper we propose a novel PANC strategy to minimize the system SPER by allocating different power levels to the network coded signals and achieve full diversity for a non-orthogonal MARC. Firstly, with the setup of IRC and IDC, respectively, we derive the decision regions for received signal and obtain the end-to-end SPER in closed form. Moreover, we propose a more efficient SPER derivation by transforming the parallelogram-shaped to the rectangle-shaped constellations, and therefore reduce the various derivations resulting from the randomization of channel conditions. By incorporating the power scaling method at relay, our proposed PANC scheme can achieve full diversity, which is proved in both theoretical and experiential aspects. In addition, to further minimize the error performance, we propose a max-min criteria based on the relationship between Euclidean distance and error probability, and formulate a convex optimization problem to obtain the optimal power adaption levels at relay. Simulations show that our method can achieve similar performances as genie-aided scheme and with low complexity. Simulation results show that the SPER derived based on our CT method can well approximate the exact SPER with a much lower complexity, and the PANC scheme with power level optimizations and power scaling factor design can achieve full diversity.

\bibliographystyle{IEEEtran}
\bibliography{IEEEabrv,PANC}

\begin{thebibliography}{10}
\providecommand{\url}[1]{#1}
\csname url@samestyle\endcsname
\providecommand{\newblock}{\relax}
\providecommand{\bibinfo}[2]{#2}
\providecommand{\BIBentrySTDinterwordspacing}{\spaceskip=0pt\relax}
\providecommand{\BIBentryALTinterwordstretchfactor}{4}
\providecommand{\BIBentryALTinterwordspacing}{\spaceskip=\fontdimen2\font plus
\BIBentryALTinterwordstretchfactor\fontdimen3\font minus
  \fontdimen4\font\relax}
\providecommand{\BIBforeignlanguage}[2]{{%
\expandafter\ifx\csname l@#1\endcsname\relax
\typeout{** WARNING: IEEEtran.bst: No hyphenation pattern has been}%
\typeout{** loaded for the language `#1'. Using the pattern for}%
\typeout{** the default language instead.}%
\else
\language=\csname l@#1\endcsname
\fi
#2}}
\providecommand{\BIBdecl}{\relax}
\BIBdecl

\bibitem{Cover:IT76}
T.~Cover and A.~Gamal, ``Capacity theorems for the relay channel,''
  \emph{{IEEE} Trans. Inform. Theory}, vol.~25, no.~5, pp. 572--584, Sep. 1979.

\bibitem{Laneman:IT04}
J.~N. Laneman, D.~N.~C. Tse, and G.~W. Wornell, ``Cooperative diversity in
  wireless networks: Efficient protocols and outage behavior,'' \emph{{IEEE}
  Trans. Inform. Theory}, vol.~50, no.~12, pp. 3062--3080, Dec. 2004.

\bibitem{Liu:book}
K.~J.~R. Liu, A.~K. Sadek, W.~Su, and A.~Kwasinski, \emph{Cooperative
  Communications and Networking}.\hskip 1em plus 0.5em minus 0.4em\relax
  Cambridge University Press, 2008.

\bibitem{Ahlswede:IT00}
R.~Ahlswede, N.~Cai, S.~Y.~R. Li, and R.~Yeung, ``Network information flow,''
  \emph{{IEEE} Trans. Inform. Theory}, vol.~46, no.~4, pp. 1204--1216, Jul.
  2000.

\bibitem{Li:IT11}
S.~Y.~R. Li, R.~W. Yeung, and N.~Cai, ``Linear network coding,'' \emph{{IEEE}
  Trans. Inform. Theory}, vol.~49, no.~2, pp. 371--381, Feb. 2003.

\bibitem{Xor:Sigcomm06}
S.~Katti, H.~Rahul, W.~Hu, D.~Katabi, M.~Medard, and J.~Crowcroft, ``Xors in
  the air: practical wireless network coding,'' in \emph{Proc. ACM SIGCOMM},
  Sep. 2006, pp. 243--254.

\bibitem{Tao:TCOM09}
T.~Cui, T.~Ho, and J.~Kliewer, ``Memoryless relay strategies for two-way relay
  channels,'' \emph{{IEEE} Trans. Commun.}, vol.~57, no.~10, pp. 3132--3143,
  Oct. 2009.

\bibitem{Zhang:JSAC09}
S.~Zhang and S.~C. Liew, ``Channel coding and decoding in a relay system
  operated with physical-layer network coding,'' \emph{{IEEE} J. Select. Areas
  Commun.}, vol.~27, no.~5, pp. 788--796, Jun. 2009.

\bibitem{Popovski:ICC06}
P.~Popovski and H.~Yomo, ``The anti-packets can increase the achievable
  throughput of a wireless multi-hop network,'' in \emph{Proc. IEEE
  International Conference on Communication (ICC 2006)}, vol.~9, Jun. 2006, pp.
  3885--3890.

\bibitem{Koike:GLOBE08}
T.~Koike-Akino, P.~Popovski, and V.~Tarokh, ``Denoising maps and constellations
  for wireless network coding in two-way relaying systems,'' in \emph{Global
  Telecommunications Conference, IEEE}, Dec. 2008, pp. 1--5.

\bibitem{Koike:JSAC09}
------, ``Optimized constellations for two-way wireless relaying with physical
  network coding,'' \emph{{IEEE} J. Select. Areas Commun.}, vol.~27, no.~5, pp.
  773--787, Jun. 2009.

\bibitem{Kramer:IT05}
G.~Kramer, M.~Gastpar, and P.~Gupta, ``Cooperative strategies and capacity
  theorems for relay networks,'' \emph{{IEEE} Trans. Inform. Theory}, vol.~51,
  no.~9, pp. 3037--3063, Sep. 2005.

\bibitem{Bao:TWC08}
X.~Bao and J.~Li, ``Adaptive network coded cooperation (ancc) for wireless
  relay networks: matching code-on-graph with network-on-graph,'' \emph{{IEEE}
  Trans. Wireless Commun.}, vol.~7, no.~2, pp. 574--583, Feb. 2008.

\bibitem{Ming:TCOM12}
M.~Xiao, J.~Kliewer, and M.~Skoglund, ``Design of network codes for
  multiple-user multiple-relay wireless networks,'' \emph{{IEEE} Trans.
  Commun.}, vol.~60, no.~12, pp. 3755--3766, Dec. 2012.

\bibitem{Li:VT12}
J.~Li, J.~Yuan, R.~Malancy, M.~Xiao, and W.~Chen, ``Full-diversity binary
  frame-wise network coding for multiple-source multiple-relay networks over
  slow-fading channels,'' \emph{{IEEE} Trans. Veh. Technol.}, vol.~61, no.~3,
  pp. 1346--1360, Mar. 2012.

\bibitem{Guan:TWC12}
W.~Guan and K.~J.~R. Liu, ``Mitigating error propagation for wireless network
  coding,'' \emph{{IEEE} Trans. Wireless Commun.}, vol.~11, no.~10, pp.
  3632--3643, Oct. 2012.

\bibitem{Jun:TWC11}
J.~Li, J.~Yuan, R.~Malaney, M.~Azmi, and M.~Xiao, ``Network coded ldpc code
  design for a multi-source relaying system,'' \emph{{IEEE} Trans. Wireless
  Commun.}, vol.~10, no.~5, pp. 1538--1551, May 2011.

\bibitem{Xiao:TCOM10}
M.~Xiao and M.~Skoglund, ``Multiple-user cooperative communications based on
  linear network coding,'' \emph{{IEEE} Trans. Commun.}, vol.~58, no.~12, pp.
  3345--3351, Dec. 2010.

\bibitem{Craig:91}
J.~Craig, ``A new, simple and exact result for calculating the probability of
  error for two-dimensional signal constellations,'' in \emph{Military
  Communications Conference, IEEE}, vol.~2, Nov. 1991, pp. 571--575.

\bibitem{Karim:WCL12}
M.~Karim, J.~Yuan, Z.~Chen, and J.~Li, ``Soft information relaying in fading
  channels,'' \emph{IEEE Wireless Commun. Lett.}, vol.~1, no.~3, pp. 233--236,
  Jun. 2012.

\bibitem{multi_user}
S.~Verdu, \emph{Multiuser Detection}.\hskip 1em plus 0.5em minus 0.4em\relax
  Cambridge University Press, 1998.

\bibitem{Peng:WCOM09}
A.~Y.~C. Peng, S.~Yousefi, and I.-M. Kim, ``On error analysis and distributed
  phase steering for wireless network coding over fading channels,''
  \emph{{IEEE} Trans. Wireless Commun.}, vol.~8, no.~11, pp. 5639--5649, Nov.
  2009.

\bibitem{Wang:TCOM08}
T.~Wang, G.~Giannakis, and R.~Wang, ``Smart regenerative relays for
  link-adaptive cooperative communications,'' \emph{{IEEE} Trans. Commun.},
  vol.~56, no.~11, pp. 1950--1960, Nov. 2008.

\bibitem{Simon:book}
M.~K. Simon and M.~S. Alouini, \emph{Digital Communication over Fading
  Channels}.\hskip 1em plus 0.5em minus 0.4em\relax {Wiley-IEEE Press}, 2004.

\bibitem{Onat:TWC08}
F.~A. Onat, A.~Adinoyi, F.~Y., H.~Yanikomeroglu, J.~S. Thompson, and M.~I. D.,
  ``Threshold selection for snr-based selective digital relaying in cooperative
  wireless networks,'' \emph{{IEEE} Trans. Wireless Commun.}, vol.~7, no.~11,
  pp. 4226--4237, Nov. 2008.

\end{thebibliography}
\begin{center}
\end{center}

\appendix
\section{Appendix}
\subsection{Computations on Wedge Probability}\label{sub:wedge}
\begin{figure}[h]
\begin{center}
\includegraphics[width=4.5in]{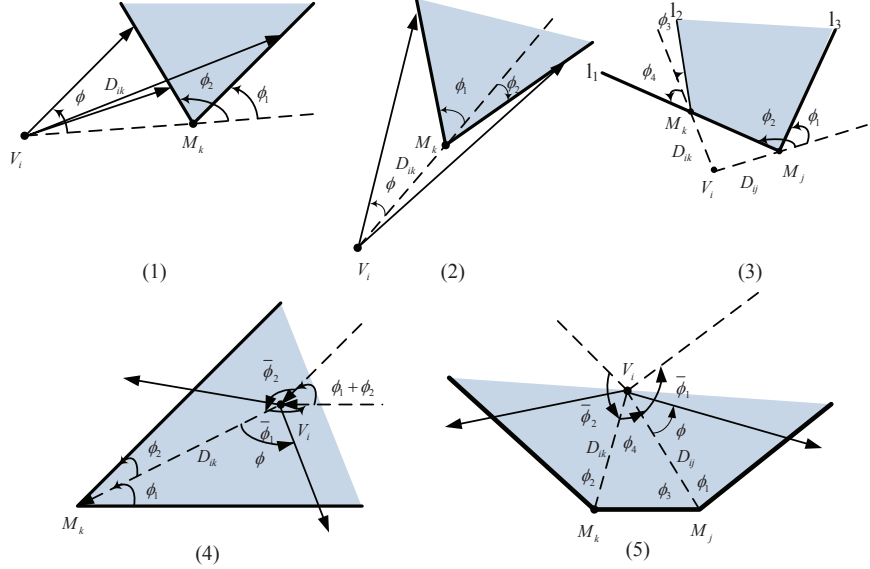}
\caption{\small{Demonstrations for the basic patterns of wedge probabilities. $d_{ik}$ (or $d_{ij}$) is the normalized distance between CP $V_i$ and wedge vertex $M_k$ (or $M_j$). In particular, both (1) and (2) are introduced in \cite{Peng:WCOM09}; (3) is the wedge difference between wedge $l_1-M_k-l_2$ and $l_1-M_j-l_3$, denoted as $l_2-M_k-M_j-l_3$, in which both $l_1$ and $l_2$ are sides of wedge $l_1-M_k-l_2$, and $l_3$ is the side of wedge $l_1-M_j-l_3$, respectively. And $\phi_i$ with $i\in\{1,\cdots,4\}$ are included angles between line $V_iM_k$ (or $V_iM_j$) and wedge side $l_m$ for $m\in\{1,2,3\}$. In (4), $\phi_i$ with $i\in\{1,2\}$ are included angle between line $V_iM_k$ and wedge sides; and $\phi_i+\bar{\phi}_i=\pi$. And in (5), $\phi_i$ with $i\in\{1,2\}$ are included angle between line $V_iM_k$ and wedge sides; and $\phi_i+\bar{\phi}_i=\pi$; $\phi_3=\angle V_iM_jM_k$ and $\phi_4=\angle M_jV_iM_k$, respectively.}}\label{fig:wedge}
\end{center}
\vspace{-5ex}
\end{figure}

Wedge probability computation method in~\cite{Craig:91,Peng:WCOM09} can be utilized to derive the SPER with irregular decision regions. In the following, we will discuss five wedge prototypes as shown in Fig.~\ref{fig:wedge}. Let us first review the wedges discussed in~\cite{Peng:WCOM09}. Denote the CP as $V_i$ and vertex of wedge as $M_k$. Assume that the angle in counter-clockwise is positive and clockwise is negative, respectively.

There are two types of wedge error probabilities to be considered when $V_i$ is outside the wedge region. For $\phi_1\phi_2 \geq 0$ as presented in Fig.~\ref{fig:wedge} (1), the wedge probability is given by~\cite{Peng:WCOM09}
\begin{equation}\label{eq:wedge1}
P_{w_1}(d_{ik},\phi_1,\phi_2)=\frac{1}{2}\left\{Q_2\left(\sqrt{2d_{ik}}\sin\phi_2;\frac{\tan^2\phi_2-1}{\tan^2\phi_2+1}\right)
-Q_2\left(\sqrt{2d_{ik}}\sin\phi_1;\frac{\tan^2\phi_1-1}{\tan^2\phi_1+1}\right)\right\},
\end{equation}
where the two-dimensional Q-function $Q_2(x;\rho)$ is defined in Notations, and its closed-form solution can be found in Equation (5.74) on~\cite{Simon:book}.

Similarly, for $\phi_1\phi_2<0$ as shown in Fig.~\ref{fig:wedge} (2), the wedge error probability is shown as~\cite{Peng:WCOM09}
\begin{equation}\label{eq:wedge2}
P_{w_2}(d_{ik},\phi_1,\phi_2)
=\frac{1}{2}\left\{Q_2\left(\sqrt{2d_{ik}}\sin\phi_1;\frac{\tan^2\phi_1-1}{\tan^2\phi_2+1}\right)
-Q_2\left(\sqrt{2d_{ik}}\sin(-\phi_2);\frac{\tan^2\phi_2-1}{\tan^2\phi_1+1}\right)\right\}.
\end{equation}

Since decision regions may also be the difference of two wedges, for notational convenience, we introduce the probability involves difference between two wedges, as shown in Fig.~\ref{fig:wedge} (3). According to the value of $\phi_1\phi_2$ and relative size of two wedges in Fig.~\ref{fig:wedge} (1) and (2), we have error probability for wedge difference given as
\begin{equation}\label{eq:wedge3}
P_{w_3}(d_{ij},d_{ik},\phi_1,\phi_2,\phi_3,\phi_4,m,n)=P_{w_m}(d_{ij},\phi_1,\phi_2)-P_{w_n}(d_{ik},\phi_3,\phi_4),
\end{equation}
where $m,n \in \{1,2\}$, angles $\phi_1$ and $\phi_2$ are with respect to vertex $M_j$, and angles $\phi_3$ and $\phi_4$ are with respect to vertex $M_k$.

Next, we discuss the probability of correct decisions, i.e., the received signal is inside the decision region of $V_i$. We use two different ways to represent the probabilities corresponding to two different decision regions, as shown in Fig.~\ref{fig:wedge} (4) and (5). In particular, the probability of a received signal within the wedge region with vertex $M_k$, shown in Fig.~\ref{fig:wedge} (4), is given by
\begin{equation}\label{eq:wedge_inside1}
\begin{aligned}
&P_{w_4}(d_{ik},\phi_1,\phi_2)=
\frac{1}{2\pi}\sum_{n=1}^2\left(1-\int_0^{\bar{\phi}_n}\exp\left(-\frac{d_{ik}\sin^2\phi_n}{\sin^2(\phi_n+\phi)}\right)d\phi\right)+\frac{\phi_1+\phi_2}{2\pi}\\
&=\frac{1}{2\pi}\sum_{n=1}^2\left(Q_2\left(\sqrt{2d_{ik}}\sin\phi_n;\frac{\tan^2\phi_n-1}{\tan^2\phi_n+1}\right)-\pi Q_1\left(\sqrt{2d_{ik}}\sin\phi_n\right)\right)+\frac{\phi_1+\phi_2+2}{2\pi}.
\end{aligned}
\end{equation}

In addition, when the decision region of $V_i$ is defined by a line segment $M_jM_k$ and two rays with initial points $M_j$ and $M_k$, respectively, shown in Fig.~\ref{fig:wedge}~(5), the probability of $V_i$ inside such wedge combination is given by
\begin{equation}\label{eq:wedge_inside2}
\begin{aligned}
P_{w_5}(d_{ik},d_{ij},\phi_1,\phi_2,\phi_3,\phi_4)
&=\frac{1}{2\pi}\left\{\sum_{n=1}^3Q_2\left(\sqrt{2d_{ij}}\sin\phi_n;\frac{\tan^2\phi_n-1}{\tan^2\phi_n+1}\right)-\sum_{n=1}^2\pi Q_1\left(\sqrt{2d_{ij}}\sin\phi_n\right)\right.\\
&\left.-Q_2\left(\sqrt{2d_{ik}}\sin(\phi_3+\phi_4);\frac{\tan^2(\phi_3+\phi_4)-1}{\tan^2(\phi_3+\phi_4)+1}\right)+\phi_1+\phi_2+\phi_4+3\right\}.\\
\end{aligned}
\end{equation}

\subsection{Derivation of the coordinate transformation at relay}\label{appendix_relay}
Define $\mathbf{Z}=[\Re\{y_{\mathcal{R}}\},\Im\{y_{\mathcal{R}}\}]^T$ as the point on the original coordinate, where $(\cdot)^T$ is the transform operation of the matrix or vector, $2 \times 1$ vector $\mathbf{Z}'$ as the intermediate transformed point, $\mathbf{A}$ as the intermediate coordinate transformation matrix. The relationship between $\mathbf{Z}$ and $\mathbf{Z}'$ is $\mathbf{Z} = \mathbf{A}\mathbf{Z}'$, where $\mathbf{A}$ is a $2 \times 2$ matrix and $\det(\mathbf{A}) \neq 0$. The probability density function of $\mathbf{Z}$ can be represented by $\mathbf{Z}'$ as
\begin{equation}\label{eq:pdf1}
\begin{aligned}
f_{\mathbf{Z}}(z)&=\frac{1}{2\pi|\Sigma|^{1/2}}\exp\left(-\frac{1}{2}(\mathbf{z}-\mathbf{V}_i)^T\Sigma^{-1}(\mathbf{z}-\mathbf{V}_i)\right)\\
&=\frac{1}{2\pi|\Sigma|^{1/2}}\exp\left(-\frac{1}{2}(\mathbf{z}'-\mathbf{A}^{-1}\mathbf{V}_i)^T\mathbf{A}^T\Sigma^{-1}\mathbf{A}(\mathbf{z}'-\mathbf{A}^{-1}\mathbf{V}_i)\right),\\
\end{aligned}
\end{equation}
where $\Sigma=[\sigma^2/2,0;0,\sigma^2/2]$, and $|\Sigma|$ is the determinant of $\Sigma$, $i\in\{1,\cdots,4\}$. Note that the covariance matrix $\mathbf{B}\triangleq\mathbf{A}^T\Sigma^{-1}\mathbf{A}$ is not diagonal, we adopt eigenvalue decomposition and obtain
\begin{equation}\label{eq:pdf2}
f_{\mathbf{Z}}(z)=\frac{1}{2\pi|\Sigma|^{1/2}}\exp\left(-\frac{1}{2}(\mathbf{z}'-\mathbf{A}^{-1}\mathbf{V}_i^T)^T[\psi_1,\psi_2]^T\left[\begin{array}{cc}\lambda_1 & 0 \\0 & \lambda_2 \\\end{array}\right][\psi_1,\psi_2](\mathbf{z}'-\mathbf{A}^{-1}\mathbf{V}_i^T)\right),\\
\end{equation}
where $\psi_i$ for $i=1,2$ and $\lambda_i$ are the eigenvectors and eigenvalues of $\mathbf{B}$, respectively. Thus, we define $\overline{\mathbf{Z}}=[\psi_1,\psi_2](\mathbf{z}'-\mathbf{A}^{-1}\mathbf{V}_i^T)$, which is a complex Gaussian random variable with mean $\overline{\mathbf{V}}_i=[\psi_1,\psi_2]\mathbf{A}^{-1}\mathbf{V}_i^T$ and covariance $[\lambda_1,0;0,\lambda_2]$, as the point on the transformed coordinate. Let $\mathbf{Q}=\left[\psi_1,\psi_2\right]$, the relationship between original received signal $\mathbf{Z}$ and the transformed received signal $\overline{\mathbf{Z}}$, original RP $\mathbf{V}_i$ and transformed RP $\overline{\mathbf{V}}_i$ after coordinate transformation and decorrelation are
\begin{equation}
\overline{\mathbf{Z}}=\mathbf{Q}\mathbf{A}^{-1}\mathbf{Z}\hspace{2mm}\text{and}\hspace{2mm}\overline{\mathbf{V}}_i=\mathbf{Q}\mathbf{A}^{-1}\mathbf{V}_i^T,
\end{equation}
respectively. To transform the RP-composed parallelogram to a rectangle centered at origin point and preserve the length of geometry's sides, e.g., $\overrightarrow{\overline{V}_i\overline{V}_j}=\overrightarrow{V_iV_j}$, the transformation matrix $\mathbf{A}$ is given by
\begin{equation}\label{eq:mn}
\mathbf{A}=\left[\begin{array}{cc}\overrightarrow{V_1V_2} & -\overrightarrow{V_1V_2} \\\overrightarrow{V_1V_3} & \overrightarrow{V_1V_3} \\\end{array}\right]
\left[\begin{array}{cc}\mathbf{V}_1(1) & \mathbf{V}_2(1) \\\mathbf{V}_1(2) & \mathbf{V}_2(2) \\\end{array}\right]^{-1}
=\frac{2}{\beta}
\left[\begin{array}{cc}|h_{1\mathcal{R}}|\Im\{h_{2\mathcal{R}}\} & -|h_{1\mathcal{R}}|\Re\{h_{2\mathcal{R}}\} \\-|h_{2\mathcal{R}}|\Im\{h_{1\mathcal{R}}\} & |h_{2\mathcal{R}}|\Re\{h_{1\mathcal{R}}\} \\\end{array}\right],
\end{equation}
where $\beta=\Re\{h_{1\mathcal{R}}\}\Im\{h_{2\mathcal{R}}\}-\Re\{h_{2\mathcal{R}}\}\Im\{h_{1\mathcal{R}}\}$. And
\begin{equation}\label{eq:B_relay}
\mathbf{B}=\frac{4}{\beta\sigma^2}\small{\left[\begin{array}{cc}
|h_{1\mathcal{R}}|^2\Im^2\{h_{2\mathcal{R}}\}+|h_{2\mathcal{R}}|^2\Im^2\{h_{1\mathcal{R}}\}
& -|h_{1\mathcal{R}}|^2\Re\{h_{2\mathcal{R}}\}\Im\{h_{2\mathcal{R}}\}-|h_{2\mathcal{R}}|^2\Re\{h_{1\mathcal{R}}\}\Im\{h_{1\mathcal{R}}\} \\-|h_{1\mathcal{R}}|^2\Re\{h_{2\mathcal{R}}\}\Im\{h_{2\mathcal{R}}\}-|h_{2\mathcal{R}}|^2\Re\{h_{1\mathcal{R}}\}\Im\{h_{1\mathcal{R}}\} & |h_{1\mathcal{R}}|^2\Re^2\{h_{2\mathcal{R}}\}+|h_{2\mathcal{R}}|^2\Re^2\{h_{1\mathcal{R}}\} \\\end{array}\right]},
\end{equation}
And for $i=1,2$, the eigenvalues are shown as
\begin{equation}\label{eq:eigenvalue}
\lambda_i=\frac{\mathbf{B}(1,1)+\mathbf{B}(2,2)\pm\sqrt{\mathbf{B}(1,1)^2+\mathbf{B}(2,2)^2+4\mathbf{B}(1,2)^2-2\mathbf{B}(1,1)\mathbf{B}(2,2)}}{2}.
\end{equation}

From \eqref{eq:B_relay} and \eqref{eq:eigenvalue} we can see that $\mathbf{B}$ is a real orthogonal symmetric matrix. In this case, matrix $\mathbf{Q}$ is a rotation matrix. Hence, the coordinate transformed by $\mathbf{A}$ is a rectangle with its sides parallel to the axis, and after eigenvalue decomposition, the new constellation is still a rectangle and being rotated counterclockwise through an angle $\theta$, which is defined by $\mathbf{Q}=[\cos(\theta),\sin(\theta);-\sin(\theta),\cos(\theta)]^T$. The corresponding eigenvectors matrix is given by \eqref{eq:Q_relay}. In this case, the final coordinate transform matrix $\mathbf{C}$ is given by
\begin{equation}
\mathbf{C}=\mathbf{Q}\mathbf{A}^{-1},
\end{equation}
where $\mathbf{A}^{-1}$ is shown in~\eqref{eq:Q_relay}.

\subsection{Proof of Theorem \ref{theorem:noalpha}}\label{sub:proposition}
Firstly, we consider the case that $T_1$ is wrongly decoded to other symbol pairs. The average probability that $T_1$ is wrongly decoded into $T_4$ at the destination is given by
\begin{equation}\label{eq:pep1}
\small
\begin{aligned}
&P(T_1 \to T_4)=\mathbb{E}\{P(T_1 \to T_4|\textbf{h})\}\\&=\mathbb{E}\left\{\sum_{k \in \{\pm a, \pm b\}}P(T_1 \to T_4|\sqrt{E_{\mathcal{R}}}x_{\mathcal{R}}=k,T_1,h_{1\mathcal{D}},h_{2\mathcal{D}},h_{\mathcal{RD}})
P(\sqrt{E_{\mathcal{R}}}x_{\mathcal{R}}=k|T_1,h_{1\mathcal{R}},h_{2\mathcal{R}})\right\}\\
&=\mathbb{E}\left\{Q_1\left[\frac{\sqrt{2}\left(\left(\sum_{i=1}^2\sqrt{E_i} |h_{i\mathcal{D}}|\right)^2+|h_{\mathcal{RD}}|^2a^2\right)}
{\sqrt{\left(E_1|h_{1\mathcal{D}}|^2 + E_2|h_{2\mathcal{D}}|^2 + |h_{\mathcal{RD}}|^2a^2\right)\sigma^2}}\right]
\left[1-\sum_{i=1}^2Q_1\left(\sqrt{2\left(E_i|h_{i\mathcal{R}}|^2/\sigma^2\right)}\right)\right.\right.\\
&\left.\left.-Q_1\left(\sqrt{2\left(\sum_{i=1}^2\sqrt{E_i} h_{i\mathcal{R}}\right)^2/\sigma^2}\right)\right]\right.
+Q_1\left[\frac{\sqrt{2}\left(\left(\sum_{i=1}^2\sqrt{E_i} |h_{i\mathcal{D}}|\right)^2+|h_{\mathcal{RD}}|^2ab\right)}
{\sqrt{\left(E_1|h_{1\mathcal{D}}|^2 + E_2|h_{2\mathcal{D}}|^2 + |h_{\mathcal{RD}}|^2a^2\right)\sigma^2}}\right]
Q_1\left(\sqrt{2\left(E_1|h_{1\mathcal{R}}|^2/\sigma^2\right)}\right)\\
&+Q_1\left[\frac{\sqrt{2}\left(\left(\sum_{i=1}^2\sqrt{E_i} |h_{i\mathcal{D}}|\right)^2-|h_{\mathcal{RD}}|^2ab\right)}
{\sqrt{\left(E_1|h_{1\mathcal{D}}|^2 + E_2|h_{2\mathcal{D}}|^2 + |h_{\mathcal{RD}}|^2a^2\right)\sigma^2}}\right]
Q_1\left(\sqrt{2\left(E_2|h_{2\mathcal{R}}|^2/\sigma^2\right)}\right)\\
&\left.+Q_1\left[\frac{\sqrt{2}\left(\left(\sum_{i=1}^2\sqrt{E_i} |h_{i\mathcal{D}}|\right)^2-|h_{\mathcal{RD}}|^2a^2\right)}
{\sqrt{\left(E_1|h_{1\mathcal{D}}|^2 + E_2|h_{2\mathcal{D}}|^2 + |h_{\mathcal{RD}}|^2a^2\right)\sigma^2}}\right]Q_1\left(\sqrt{2\left(\sum_{i=1}^2\sqrt{E_i} h_{i\mathcal{R}}\right)^2/\sigma^2}\right)\right\}.
\end{aligned}
\end{equation}

Without loss of generality, we assume $E_1=E_2=E_{\mathcal{R}}=E$ in the following. Denote $\rho=E/\sigma^2$ as the reference system SNR for $i\in\{1,2,\mathcal{R}\}$. In general, averaging the following one-dimensional Q-function over channel distributions, we have
\begin{equation}\label{eq:channave}
\mathbb{E}\left\{Q_1\left(\sqrt{2\rho|h_{ij}|^2}\right)\right\}=\frac{1}{\pi}\int_{0}^{\pi/2}\left(1+\frac{\rho\gamma_{ij}}{\sin^2\theta}\right)^{-1}d\theta\mathop  \approx \limits^{\rho \rightarrow \infty}\frac{1}{4\gamma_{ij}}\rho^{-1}.
\end{equation}
Likewise, $\mathbb{E}\left\{Q_1\left(\sqrt{2\rho\sum_{t\in\{ij,mn\}}|h_t|^2}\right)\right\}\mathop  \approx \limits^{\rho \rightarrow \infty}\frac{3}{16\gamma_{ij}\gamma_{mn}}\rho^{-2}$ and $\mathbb{E}\{Q_1(\sqrt{2\rho\sum_{t\in\{ij,mn,pq\}}|h_t|^2})\}$\\$\mathop  \approx \limits^{\rho \rightarrow \infty}\frac{5}{32\gamma_{ij}\gamma_{mn}\gamma_{pq}}\rho^{-3}$. According to the result shown in \cite{Onat:TWC08}, we have the high-SNR approximation
\begin{equation}\label{eq:approx}
\mathbb{E}\left\{Q_1\left[\frac{\sqrt{2}\left(\left(\sum_{i=1}^2\sqrt{E_i} |h_{i\mathcal{D}}|\right)^2-|h_{\mathcal{RD}}|^2ab\right)}
{\sqrt{\left(E_1|h_{1\mathcal{D}}|^2 + E_2|h_{2\mathcal{D}}|^2 + |h_{\mathcal{RD}}|^2a^2\right)\sigma^2}}\right]\right\}
\approx \frac{\gamma_{\mathcal{RD}}}{\gamma_{1\mathcal{D}}+\gamma_{2\mathcal{D}}+\gamma_{\mathcal{RD}}}.
\end{equation}

With the conclusion in \eqref{eq:channave} and \eqref{eq:approx}, averaging the probability $P(T_1 \to T_4|\textbf{h})$ in \eqref{eq:pep1} over channel distributions, we further have
\begin{equation}\label{eq:pep1app}
\begin{aligned}
P(T_1 \to T_4)&\approx \frac{5}{32\gamma_{1\mathcal{D}}\gamma_{2\mathcal{D}}\gamma_{\mathcal{RD}}}\rho^{-3}
+\frac{5}{128\gamma_{1\mathcal{D}}\gamma_{2\mathcal{D}}\gamma_{\mathcal{RD}}\gamma_{1\mathcal{R}}}\rho^{-4}\\
&+\frac{\gamma_{\mathcal{RD}}}{4\gamma_{2\mathcal{R}}(\gamma_{1\mathcal{D}}+\gamma_{2\mathcal{D}}+\gamma_{\mathcal{RD}})}\rho^{-1}
+\frac{\gamma_{\mathcal{RD}}}{4\gamma_{\mathcal{SR}}(\gamma_{1\mathcal{D}}+\gamma_{2\mathcal{D}}+\gamma_{\mathcal{RD}})}\rho^{-1},
\end{aligned}
\end{equation}
where $\gamma_{\mathcal{SR}}=\gamma_{1\mathcal{R}}+\gamma_{2\mathcal{R}}$. Likewise, we can derive $P(T_2 \to T_3)$. When only one error occurs, we further have
\begin{equation}\label{eq:pep3}
P(T_1 \to T_2)\approx
\left\{
\begin{array}{cc}
\frac{3}{16\gamma_{1\mathcal{D}}\gamma_{\mathcal{RD}}}\rho^{-2}
+\frac{1}{4\gamma_{1\mathcal{R}}}\rho^{-1}
+\frac{1}{4\gamma_{2\mathcal{R}}}\rho^{-1}
+\frac{1}{4\gamma_{\mathcal{SR}}}\rho^{-1}, & \text{when}\hspace{2mm} a>b,\\
\frac{3}{16\gamma_{1\mathcal{D}}\gamma_{\mathcal{RD}}}\rho^{-2}
+\frac{1}{4\gamma_{1\mathcal{R}}}\rho^{-1}
+\frac{3}{64\gamma_{1\mathcal{D}}\gamma_{\mathcal{RD}}\gamma_{2\mathcal{R}}}\rho^{-3}
+\frac{3}{64\gamma_{1\mathcal{D}}\gamma_{\mathcal{RD}}\gamma_{\mathcal{SR}}}\rho^{-3}, & \text{when}\hspace{2mm} a<b,\\
\frac{3}{16\gamma_{1\mathcal{D}}\gamma_{\mathcal{RD}}}\rho^{-2}
+\frac{1}{4\gamma_{1\mathcal{R}}}\rho^{-1}
+\frac{1}{16\gamma_{1\mathcal{D}}\gamma_{2\mathcal{R}}}\rho^{-2}
+\frac{1}{16\gamma_{1\mathcal{D}}\gamma_{\mathcal{SR}}}\rho^{-2}, & \text{when}\hspace{2mm} a=b. \\
\end{array}
\right.
\end{equation}
Similarly, we can derive $P(T_1 \to T_3)$, $P(T_2 \to T_4)$.
From \eqref{eq:pep1app}, and \eqref{eq:pep3}, we can conclude that the PANC scheme can only achieve one order diversity in MARC system without power scaling.

In NC based MARC system, averaging the related PEPs over the channel coefficients, we have
\begin{equation}
\begin{aligned}
P(T_1 \to T_4)& \approx \frac{5}{32\gamma_{\mathcal{SD}}\gamma_{\mathcal{RD}}\gamma_{1\mathcal{R}}}\rho^{-3}
+\frac{5}{32\gamma_{\mathcal{SD}}\gamma_{\mathcal{RD}}\gamma_{2\mathcal{R}}}\rho^{-3}
+\frac{1}{4\gamma_{\mathcal{SD}}}\rho^{-1},\\
P(T_1 \to T_2)& \approx \frac{3}{16\gamma_{1\mathcal{D}}\gamma_{1\mathcal{R}}}\rho^{-2}
+\frac{3}{16\gamma_{1\mathcal{D}}\gamma_{2\mathcal{R}}}\rho^{-2}
+\frac{3}{16\gamma_{1\mathcal{D}}\gamma_{\mathcal{RD}}}\rho^{-2}.
\end{aligned}
\end{equation}
Similarly, we can derive $P(T_1 \to T_3)$, $P(T_2 \to T_3)$ and $P(T_2 \to T_4)$. In this case, we can conclude that the MARC system applied network coding cannot achieve full diversity.

\subsection{Proof of Theorem \ref{tho:theorem1}}\label{sub:Diversity_NCPA}
We prove Theorem~2 using our virtual channel model. After some manipulations, it is easy to show that given any power scaling coefficient $\alpha$ employed at the relay side, the lower bound of SPER can be in general approximated as
\begin{equation}\label{eq:pep}
\begin{aligned}
&P_v\triangleq P \left( {(x_1,x_2,x_{\mathcal{R}}) \to (\hat{x}_1,\hat{x}_2,\hat{x}_{\mathcal{R}})} \right)\\
\approx &\mathbb{E}\left[ {Q\left( {\frac{{{{\left( \sqrt{E_1}|{h_{1\mathcal{D}}|\left( {{x_1} - {{\hat x}_1}} \right) + \sqrt{E_2}|h_{2\mathcal{D}}|\left( {{x_2} - {{\hat x}_2}}
\right)} \right)}^2} + \alpha|h_{\mathcal{RD}}|^2(x_{\mathcal{R}}-\hat{x}_{\mathcal{R}})^2 }}
{{\sqrt {E_1|h_{1\mathcal{D}}|^2(x_1-\hat{x}_1)^2 + E_2|h_{2\mathcal{D}}|^2(x_2-\hat{x}_2)^2 + \alpha|h_{\mathcal{RD}}|^2(x_{\mathcal{R}}-\hat{x}_{\mathcal{R}})^2 } }}} \right)} \right]\\
\mathop\le\limits^{{{(x + y)}^2} \le 2\left( {{x^2} + {y^2}} \right)}&\mathbb{E}\left[ {Q\left( {\frac{{{{\left( \sqrt{E_1}|{h_{1\mathcal{D}}|\left( {{x_1} - {{\hat x}_1}} \right) + \sqrt{E_2}|h_{2\mathcal{D}}|\left( {{x_2} - {{\hat x}_2}}
\right)} \right)}^2} + \alpha|h_{\mathcal{RD}}|^2(x_{\mathcal{R}}-\hat{x}_{\mathcal{R}})^2 }}
{{2\sqrt {\left(\sqrt{E_1}|h_{1\mathcal{D}}|(x_1-\hat{x}_1) + \sqrt{E_2}|h_{2\mathcal{D}}|(x_2-\hat{x}_2)\right)^2 + \alpha|h_{\mathcal{RD}}|^2(x_{\mathcal{R}}-\hat{x}_{\mathcal{R}})^2 } }}} \right)} \right].
\end{aligned}
\end{equation}
After applying the Chernoff bound $Q_1(x)\leq \frac{1}{2}\exp\left(-\frac{x^2}{2}\right)$, we can further obtain
\begin{equation}\label{eq:pepbound}
\small
P_v\leq \mathbb{E}\left[\frac{1}{2}\exp\left(-\frac{\left(\sqrt{E_1}|h_{1\mathcal{D}}|(x_1-\hat{x}_1)+\sqrt{E_2}|h_{2\mathcal{D}}|(x_2-\hat{x}_2)\right)^2
+\gamma_{\mathcal{SRD}}|x_{\mathcal{R}}-\hat{x}_\mathcal{R}|^2}{4}\right)\right]
\mathop \approx \limits^{\rho \to \infty}\frac{1}{2}{\left( {\prod\limits_{k = 1}^r {{\Lambda _i}} } \right)^{ - 1}}{\rho ^{ - r}}
\end{equation}
where $\Lambda_i$, $r$ are the $i$th non-zero eigenvalue and the rank of the diagonal matrix\\ $[
\frac{(\sqrt{E_1}|h_{1\mathcal{D}}|(x_1-\hat{x}_1)+\sqrt{E_2}|h_{2\mathcal{D}}|(x_2-\hat{x}_2))^2}{4},0;0,{\frac{{{\gamma_{\mathcal{SRD}}}{{( x_{\mathcal{R}}-\hat{x}_{\mathcal{R}} )}^2}}}{4}}]$, respectively. $\gamma_{\mathcal{SRD}}=\min\{\gamma_{\mathcal{SR}},\gamma_{\mathcal{RD}}\}$ is an exponential distributed random variable with mean $\frac{\gamma_{1\mathcal{R}}\gamma_{2\mathcal{R}}\gamma_{\mathcal{RD}}}{\gamma_{1\mathcal{R}}\gamma_{2\mathcal{R}}
+\gamma_{1\mathcal{R}}\gamma_{\mathcal{RD}}+\gamma_{2\mathcal{R}}\gamma_{\mathcal{RD}}}$, which is proven as follows. Define $T=\min ({E_1|h_{1\mathcal{R}}|^2},{E_2|h_{2\mathcal{R}}|^2})$. Since $\gamma_{\mathcal{SRD}}
=\min\{E_1|h_{1\mathcal{R}}|^2,E_2|h_{2\mathcal{R}}|^2,(\sqrt{E_1}|h_{1\mathcal{R}}|+\sqrt{E_2}|h_{2\mathcal{R}}|)^2,\gamma_{\mathcal{RD}}\}$, we have
\begin{equation}
2\gamma_{\mathcal{SRD}}\geq \min\{2E_1|h_{1\mathcal{R}}|^2,2E_2|h_{2\mathcal{R}}|^2,E_1|h_{1\mathcal{R}}|^2+E_2|h_{2\mathcal{R}}|^2,2\gamma_{\mathcal{RD}}\}
\geq 2\min\{T,\gamma_{\mathcal{RD}}\}.
\end{equation}
If $X$ and $Y$ are i.i.d. exponentially distributed random variables with mean $v_x$ and $v_y$, respectively, $\min\{X,Y\}$ is a also an exponentially distributed random variable with mean $v_x+v_y$. So we have the conclusion of $\gamma_{\mathcal{SRD}}$. Since both two diagonal elements are non-zero when an error event happens or two error events happen, we have $\mathop {\max }\limits_{\hat x \ne x} \Pr \left( {x \to \hat x} \right) = {\rm O}\left( {{\Gamma ^{ - 2}}} \right)$. Note that, replacing $\gamma_{\mathcal{RD}}$ in \eqref{eq:pepbound} by $\bar{\gamma}_{\mathcal{RD}}$ will not change the result of diversity, which completes the proof of the theorem.

If we apply the power scaling on CXNC scheme, when one error occurs, i.e., $\hat{x}_i=-x_i$ for $i\in\{1,2\}$, we have
\begin{equation}\label{eq:cxnc1}
\begin{aligned}
&P \left( {(x_1,x_2,x_{\mathcal{R}}) \to (-\hat{x}_1,\hat{x}_2,\hat{x}_{\mathcal{R}})} \right)\\
&\approx \mathbb{E}\left[ {Q\left( {\frac{{{{ 2E_1|{h_{1\mathcal{D}}|^2} }} }}
{{\sqrt {E_1|h_{1\mathcal{D}}|^2(x_1-\hat{x}_1)^2 + E_2|h_{2\mathcal{D}}|^2(x_2-\hat{x}_2)^2 + \alpha|h_{\mathcal{RD}}|^2(x_{\mathcal{R}}-\hat{x}_{\mathcal{R}})^2 } }}} \right)} \right].\\
\end{aligned}
\end{equation}
Averaging the probability in \eqref{eq:cxnc1} over channel distributions, we further have
\begin{equation}\label{eq:cxnc2}
P \left( {(x_1,x_2,x_{\mathcal{R}}) \to (-\hat{x}_1,\hat{x}_2,\hat{x}_{\mathcal{R}})} \right)\approx \frac{1}{4\gamma_{1\mathcal{D}}}\rho^{-1}.
\end{equation}
Hence, we can conclude that the power scaled CXNC scheme still cannot achieve full diversity.

\end{document}